\newcommand*{\ATLASLATEXPATH}{}
\documentclass[PAPER, cernpreprint, atlasdraft=false, texlive=2016, UKenglish]{\ATLASLATEXPATH atlasdoc}
 
\usepackage{\ATLASLATEXPATH atlaspackage}
\usepackage{\ATLASLATEXPATH atlasbiblatex}
 
\usepackage{\ATLASLATEXPATH atlasphysics}
\usepackage{multirow}
\usepackage{dcolumn}
 
\graphicspath{{logos/}{figures/}}
 
\usepackage{ANA-HIGG-2017-08-PAPER-defs}

\AtlasTitle{Searches for lepton-flavour-violating decays of the Higgs boson
in $\sqrt{s}=13$\,\TeV\ $pp$ collisions with the ATLAS detector}

\AtlasAbstract{
This Letter presents direct searches for lepton flavour violation in Higgs boson decays,
$H\rightarrow e\tau$ and $H\rightarrow\mu\tau$, performed with the ATLAS detector at the LHC. The searches are based on a data sample of proton--proton
collisions
at a centre-of-mass energy $\sqrt{s} = 13$\,\TeV, corresponding to
an integrated luminosity of $36.1\,\mathrm{fb}^{-1}$.
No significant excess is observed above the expected background from Standard Model processes.
The observed (median expected) 95\% confidence-level upper limits on the lepton-flavour-violating branching ratios are
$0.47\%$ ($0.34^{+0.13}_{-0.10}\,\%$) and $0.28\%$
($0.37^{+0.14}_{-0.10}\,\%$) for $H\to e\tau$ and $H\to\mu\tau$,
respectively.
}

\AtlasRefCode{HIGG-2017-08}
 
\PreprintIdNumber{CERN-EP-2019-126}
 
\AtlasDate{\today}

\arXivId{1907.06131}

\AtlasJournalRef{Phys. Lett. B 800 (2020) 135069}
\AtlasDOI{10.1016/j.physletb.2019.135069}

\AtlasCoverSupportingNote{Search for lepton-flavour-violating decays of the Higgs boson in 13 TeV collisions with the ATLAS Detector}{https://cds.cern.ch/record/2261629}
\AtlasCoverSupportingNote{Measurement of the $H\rightarrow\tau^{+}\tau^{-}$ cross-section in 13\,TeV Collisions with the ATLAS Detector}{https://cds.cern.ch/record/2261605}

\AtlasCoverAnalysisTeam{Attilio Andreazza, Mohamad Kassem Ayoub, Swagato Banerjee, Elisabetta Luigia Barberio, Fabian Horst Becherer, Diptaparna Biswas, Ulrike Blumenschein, Quentin Buat, Vincenzo Cavasinni, Xin Chen, Elias Coniavitis, Vincent Croft, Nguyen Phuong Dang, Tomas Davidek, Antonio De Maria, Pier-Olivier Deviveiros, Eric Drechsler, Luca Fiorini, Christian Grefe, Minoru Hirose, Julia Mariana Iturbe Ponce, Michel Janus, Karsten Koeneke, Brian Le, Alena Loesle, Luis Flores Castillo, Theo Jean Megy, Michaela Mlynarikova, Alessia Murrone, Clara Nellist, Carlo Enrico Pandini, Atanu Pathak, Arnulf Quadt, Elzbieta Richter-Was, Sergi Rodriguez Bosca, Dirk Sammel, Frank Sauerburger, Daniel Scheirich, Katharina Schleicher, Ulrike Schnoor, Markus Schumacher, Federico Scutti, Duc Bao Ta, Soshi Tsuno, Ligang Xia, Daniele Zanzi, Lei Zhang, Zhidong Zhang, Zinonas Zinonos}
 
\AtlasCoverEdBoardMember{Borut Paul Kersevan~(chair), Daniel Froidevaux, Evelin Meoni, Stefania Xella}
 
\AtlasCoverEgroupEditors{atlas-higg-2017-08-editors@cern.ch}
 
\AtlasCoverEgroupEdBoard{atlas-higg-2017-08-editorial-board@cern.ch}
\hypersetup{pdftitle={ATLAS document},pdfauthor={The ATLAS Collaboration}}
 
\begin{document}
 
\maketitle

\section{Introduction}
The search for processes beyond the Standard Model (SM) is one of the
main goals of the Large Hadron Collider (LHC) programme at CERN.
A possible sign of such processes is lepton flavour violation (LFV) in decays of the Higgs boson~\cite{HIGG-2012-27,CMS-HIG-12-028}.
Many beyond-SM theories predict LFV decays of the Higgs boson, such as supersymmetry~\cite{Arana-Catania:2013xma,Arhrib:2012ax}, other models with more than one Higgs doublet~\cite{Bjorken:1977vt,DiazCruz:1999xe}, composite Higgs models~\cite{Agashe:2009di}, models with flavour symmetries~\cite{Ishimori:2010au} or warped extra dimensions~\cite{Perez:2008ee,Azatov:2009na,Albrecht:2009xr} models and others~\cite{Goudelis:2011un,McKeen:2012av}.
 
In this Letter, searches for LFV decays of the Higgs boson, $H\rightarrow e\tau$ and
$H\rightarrow\mu\tau$, at the LHC with the ATLAS experiment are presented.
Studies are based on proton--proton ($pp$) collision data recorded in 2015--2016 at a centre-of-mass energy $\sqrt{s}=13\,\TeV$.
The dataset corresponds to an integrated luminosity of $36.1\,\ifb$.
 
Previous ATLAS searches~\cite{HIGG-2014-08,HIGG-2015-09} placed an upper limit of 1.04\% (1.43\%) on the
$H\rightarrow e\tau$ ($H\rightarrow \mu\tau$) branching ratio ($\mathcal{B}$) with a 95\%
confidence level (CL) using Run~1 data collected at $\sqrt{s}=8\,\TeV$, corresponding to an integrated luminosity of $20.3\,\ifb$.
The CMS Collaboration recently provided 95\% CL
upper limits on these branching ratios of 0.61\% and 0.25\%, respectively,
using data collected at $\sqrt{s}=13\,\TeV$, with an integrated luminosity
of $35.9\,\ifb$~\cite{CMS-HIG-17-001}.
 
The searches presented here involve both leptonic ($\tau\to\ell^\prime\nu\bar{\nu}$\,\footnote{Unless explicitly mentioned otherwise, leptons (denoted by
$\ell$ or $\ell^\prime$) refer to electrons or muons.}) and hadronic
($\tau\to\mathrm{hadrons}+\nu$) decays of $\tau$-leptons, denoted {\tlep} and {\thad} respectively.
The dilepton final state {\leplep} only considers pairs of different-flavour leptons. Same-flavour lepton pairs are rejected due to the large lepton pair-production Drell-Yan background.
Two channels are considered for each of the two searches:
$e\tau_{\mu}$ and $e\thad$ for the $H\to e\tau$ search, $\mu\tau_{e}$ and $\mu\thad$ for the $H\to\mu\tau$ search.
The analysis is designed such that any potential LFV signal overlap between the $H\to e\tau$ and $H\to\mu\tau$ searches is negligible.
Many methods are reused from the measurement of the Higgs boson cross-section
in the $H\to\tau\tau$ final state~\cite{htautau}.
 
The ATLAS detector\footnote{ATLAS uses a right-handed coordinate system with
its origin at the nominal interaction point in the centre of the detector
and the $z$-axis along the beam pipe. The azimuthal angle $\phi$ runs around
the beam pipe, the pseudorapidity is defined in terms of the polar angle
$\theta$ as $\eta\equiv-\ln\tan(\theta/2)$. Angular distance in the $\eta$--$\phi$
space is defined as $\Delta R\equiv\sqrt{(\Delta\eta)^2+(\Delta\phi)^2}$.}
is described in
Refs.~\cite{PERF-2007-01,ATLAS-TDR-19,PIX-2018-001}. It consists of an inner tracking detector covering the range $|\eta| < 2.5$, surrounded by a superconducting solenoid
providing a 2\,T axial magnetic field, high-granularity electromagnetic
($|\eta|<3.2$) and hadronic calorimeters ($|\eta|<4.9$), and a muon spectrometer (MS) which covers the range $|\eta| < 2.7$ and includes fast trigger chambers ($|\eta| < 2.4$) and superconducting toroidal magnets.
\section{Simulation samples}
Samples of Monte Carlo (MC) simulated events are used to optimize the event
selection, and to model the signal and several of the background processes.
The samples were
produced with the ATLAS simulation infrastructure~\cite{ATLASSIM} using
the full detector simulation performed by the {\GEANT}4~\cite{Geant4} toolkit.
The Higgs boson mass was set to  $m_{H}=125\,\GeV$~\cite{HIGG-2014-14}.
The four leading Higgs boson production mechanisms are considered: the
gluon--gluon fusion (ggF), vector-boson fusion (VBF) and two associated
production modes ($WH$, $ZH$), while the others give negligible contributions and are ignored.
The cross-sections of all Higgs boson production processes were normalized to the SM predictions~\cite{LHCXS}.
The LFV Higgs boson decays as well as the $H\rightarrow\tau\tau$ and $H\rightarrow WW$ background decays were modelled with Pythia~8~\cite{Sjostrand:2014zea}.
Other background processes involve electroweak production
of $W/Z$ bosons via VBF, Drell--Yan production of $W/Z$ in
association with jet(s) as well as diboson, single top-quark and top-quark pair ($t\bar{t}$) production. The MC generators used for the SM $H\to\tau\tau$ cross-section measurement~\cite{htautau} were also employed here for all background components. The generators and parton shower models used to simulate different processes are summarized in Table~\ref{tab:generators}.
 
\begin{table}[htb]
\caption{Generators used to describe the signal and background processes, parton distribution function (PDF) sets for the hard process, and models used for parton showering, hadronization and the underlying event (UEPS). The orders of the total cross-sections used to normalize the events are also given. More details are given in Ref.~\cite{htautau}.}
\centering
\setlength{\extrarowheight}{0pt}
\scriptsize
\begin{tabular}{lllll}
\toprule
Process & Generator & PDF & UEPS & Cross-section order \\ \midrule
ggF & Powheg-Box v2 \cite{Nason:2004rx,Powheg_2,Powheg_3,Powheg_4,Powheg_6} NNLOPS \cite{Powheg_5} & PDF4LHC15 \cite{Butterworth:2015oua} NNLO & Pythia 8.212 \cite{Sjostrand:2014zea} & N$^3$LO QCD + NLO EW \cite{ggH_xs_1,ggH_xs_2,ggH_xs_3,ggH_xs_4} \\
VBF & Powheg-Box v2 MiNLO \cite{Powheg_6} & PDF4LHC15 NLO & Pythia 8.212 & $\sim$NNLO QCD + NLO EW \cite{vbf_xs_1,vbf_xs_2,vbf_xs_3} \\
$WH$, $ZH$ & Powheg-Box v2 MiNLO & PDF4LHC15 NLO & Pythia 8.212 & NNLO QCD + NLO EW \cite{vh_xs_1,vh_xs_2,vh_xs_3} \\ \midrule
$W$/$Z$+jets & Sherpa 2.2.1 \cite{Gleisberg:2008ta} & NNPDF30NNLO \cite{Ball:2014uwa} & Sherpa 2.2.1 \cite{Sherpa_UEPS} & NNLO \cite{Melnikov:2006kv,Anastasiou:2003ds} \\
$VV$/$V\gamma^*$ & Sherpa 2.2.1 & NNPDF30NNLO & Sherpa 2.2.1 & NNLO \\
$t\bar{t}$ & Powheg-Box v2 \cite{Nason:2004rx,Powheg_2,Powheg_3,ttbar_1} & CT10 \cite{CT10} & Pythia 6.428 \cite{Sjostrand:2006za} & NNLO+NNLL \cite{ttbar_xs}\\
Single $t$ & Powheg-Box v1 \cite{Powheg_7,Powheg_8} & CT10 & Pythia 6.428 & NLO \cite{single_t_xs_1,single_t_xs_2,single_t_xs_3} \\ \bottomrule
\end{tabular}
\label{tab:generators}
\end{table}
\section{Object reconstruction}
The correct identification of $H\rightarrow\ell\tau$ events requires reconstruction of
several different objects (electrons, muons, and jets, including those
initiated by hadronic decays of $\tau$-leptons) and the missing
transverse momentum $\vec{p}_{\mathrm{T}}^{\mathrm{miss}}$, whose magnitude is called $\MET$.
 
Electrons are reconstructed by matching tracks in the inner detector to clustered energy
deposits in the electromagnetic calorimeter~\cite{PERF-2017-03}. Loose likelihood-based
identification~\cite{PERF-2017-01},
$\pT > 15\,\GeV$ and fiducial volume requirements ($|\eta| < 2.47$, excluding
the transition region between the barrel and the endcap calorimeters $1.37 < |\eta| < 1.52$) are applied.
Medium identification, corresponding to an efficiency of 87\% at $\pt=20\,\GeV$, 
is imposed for the baseline electron selection.
 
Muons are identified by tracks reconstructed in the inner detector and matched to tracks
in the MS. Loose identification~\cite{PERF-2015-10}, $\pT > 10\,\GeV$ and
$|\eta| < 2.5$ requirements are applied. Medium
identification (efficiency of 96.1\% for muons with $\pt>20\,\GeV$) 
is imposed for the baseline muon selection.
 
Isolation criteria exploiting calorimeter and track-based information are applied to both electrons and muons. The gradient working point is used, featuring an efficiency of 90\% (99\%) obtained for leptons with $\pt>25\,\GeV$ ($60\,\GeV$) originating from the $Z\to\ell\ell$ process~\cite{PERF-2017-01,PERF-2015-10}.
 
Jets are reconstructed using the anti-$k_{t}$ algorithm~\cite{Cacciari:2008gp} as implemented by the FastJet~\cite{Fastjet} package. The algorithm is applied to topological clusters of calorimeter cells~\cite{PERF-2014-07} with a radius parameter $R=0.4$.
Only jets with $\pT > 20\,\GeV$ and $|\eta| < 4.5$ are considered. Jets from other $pp$ interactions in the same and neighbouring bunch crossings (pile-up) are suppressed using jet vertex tagger (JVT) algorithms~\cite{ATLAS-CONF-2014-018,PERF-2016-06}.
Jets containing $b$-hadrons ($b$-jets) are identified by the MV2c20 algorithm~\cite{PERF-2012-04,ATL-PHYS-PUB-2016-012} in the central region ($|\eta| < 2.4$).
A working point corresponding to 85\% average efficiency determined for $b$-jets
in $t\bar{t}$ simulated events is chosen, rejection factors are 2.8 and 28 against $c$-jets and light-flavour jets respectively.
 
The reconstruction of the object formed by the visible products of the
$\thad$ decay (\tauhadvis) begins from jets reconstructed by the anti-$k_{t}$ jet algorithm with a
radius parameter $R=0.4$. Information from the inner detector tracks associated with the energy deposits in the calorimeter is incorporated in the reconstruction.
Only {\tauhadvis} candidates with $\pT > 20\,\GeV$
and $|\eta| < 2.5$ are considered.\footnote{The transition region in
$\eta$ is excluded, similarly to electrons.}
One or three associated tracks with an absolute total charge $|q|=1$ are
required. An identification algorithm~\cite{ATL-PHYS-PUB-2015-045,ATLAS-CONF-2017-029} based on boosted
decision trees (BDT)~\cite{bdt,bdt:adaboost,bdt:gradboost} is used to
reject {\tauhadvis} candidates arising from misidentification of jets or from
decays of hadrons with $b$- or $c$-quark content.
Unless otherwise indicated, a tight identification (ID) working point is used for
the {\tauhadvis}, corresponding to an efficiency of 60\%
(45\%) for 1-prong (3-prong) candidates.
Jets corresponding to identified {\tauhadvis} candidates are removed from the jet collection.
The {\tauhadvis} candidates with one track overlapping with
an electron candidate with high ID score, as
determined by a  multivariate (MVA) approach, are rejected. Leptonic $\tau$-decays are reconstructed as electrons or muons.
 
Events considered in the analysis are triggered with single-electron or single-muon triggers. The {\pt} thresholds depend on the isolation requirement and data-taking period~\cite{TRIG-2016-01,ATL-DAQ-PUB-2017-001}. The lowest trigger thresholds correspond to $25-27\,\GeV$ (electrons) and $21-27\,\GeV$ (muons).
 
\section{Event selection and categorization}
\label{sec:event_selection}
 
\begin{table}
\caption{Baseline event selection and further categorization for
the {\leplep} and {\lephad} channels.
The same criteria are also used for the control region (CR)
definitions in the {\leplep} channel
(Section~\ref{sec:background}), but one requirement of the
baseline selection is inverted to achieve orthogonal event selection.
There is no CR in the {\lephad} channel.
}
\centering
\small{
\setlength{\extrarowheight}{2pt}
\begin{tabular}{lcc}
\toprule
Selection & \leplep & \lephad \\
\midrule
\multirow{6}{*}{Baseline}
& exactly $1e$ and $1\mu$, OS & exactly $1\ell$ and
$1\tauhadvis$, OS \\
& $\pt^{\ell_1} > 45\,\GeV$ & $\pt^{\ell} > 27.3\,\GeV$\\
& $\pt^{\ell_2} > 15\,\GeV$ & $\pt^{\tauhadvis} > 25\,\GeV$, $|\eta^{\tauhadvis}| < 2.4$ \\
& $30\,\GeV < \mvis < 150\,\GeV$ & $\sum\limits_{i=\ell,\tauhadvis}\!\!\!\!\!\!\!\!\cos\Delta\phi(i,\MET) > -0.35$ \\
& $\pt^{e}(\text{track})/\pt^{e}(\text{cluster}) < 1.2$ ($\mu\tau_{e}$ only) &
$|\Delta\eta(\ell,\tauhadvis)|<2$ \\
& \multicolumn{2}{c}{$b$-veto (for jets with $\pt > 25\,\GeV$ and
$|\eta| < 2.4$)} \\
\midrule
\multirow{4}{*}{VBF} & \multicolumn{2}{c}{Baseline} \\
& \multicolumn{2}{c}{$\ge 2$ jets, $\pt^{\mathrm{j}_1} > 40\,\GeV$,
$\pt^{\mathrm{j}_2}>30\,\GeV$} \\
& \multicolumn{2}{c}{$|\detajj| > 3$, $\mjj > 400\,\GeV$}\\
& $-$ & $\pt^{\tauhadvis} > 45\,\GeV$ \\
\midrule
\multirow{5}{*}{Non-VBF} & \multicolumn{2}{c}{Baseline plus fail VBF categorization} \\
& $\mt(\ell_1,\MET) > 50\,\GeV$ & $-$ \\
& $\mt(\ell_2,\MET) < 40\,\GeV$ & $-$\\
& $|\Delta\phi(\ell_2,\MET)| < 1.0$ & $-$\\
& $\pt^{\tau}/\pt^{\ell_1} > 0.5$ & $-$\\
\midrule
Top-quark CR & inverted $b$-veto: &   \\
VBF and non-VBF & $\ge 1\ b$-tagged jet ($\pt>25\,\GeV$ and $|\eta|<2.4$) &  \\
\midrule
$Z\to\tau\tau$ CR & inverted $\pt^{\ell_1}$ requirement: & \\
VBF and non-VBF & $35\,\GeV < \pt^{\ell_1} < 45\,\GeV$ & \\
\bottomrule
\end{tabular}
}
\label{tab:event_selection}
\end{table}
Events selected in the {\leplep} channel contain exactly one
electron and one muon of opposite-sign (OS) charges. Similarly in the
{\lephad} channel, a lepton and a {\tauhadvis} of OS
charges are required, and events with more than one baseline lepton are rejected. The selection criteria
are summarized in Table~\ref{tab:event_selection} for the analysis
categories as well as the control regions (CRs),
which are described in Section~\ref{sec:background}.

In the {\leplep} channel, $\ell_1$ and $\ell_2$ denote the leading and
subleading lepton in {\pt}, respectively. Events where the leading lepton is an electron (muon) are used in the search for $H\to e\tau_{\mu}$ ($H\to\mu\tau_{e}$). A requirement on the
dilepton invariant mass, equal to the invariant mass of the lepton
and the visible $\tau$-decay products, {\mvis}, reduces backgrounds
with top quarks, and the criterion applied to the track-to-cluster {\pt} ratio of the electron reduces the $Z\to\mu\mu$ background where a muon deposits a large amount of energy in the electromagnetic calorimeter and is misidentified as an electron in the $\mu\tau_{e}$ channel. The contribution from the $H\to\tau\tau$ decay is reduced by the asymmetric {\pt} selection of the two leptons.
 
In the {\lephad} channel, the criterion based on the
azimuthal separations of lepton--{\MET} and {\tauhadvis}--{\MET},
$\sum_{i=\ell,\tauhadvis}\cos\Delta\phi(i,\MET)$,
reduces the $W+$jets background whereas the requirement on
$|\Delta\eta(\ell,\tauhadvis)|$ reduces backgrounds
with misidentified {\tauhadvis} candidates.
 
For both channels of each search, a $b$-veto requirement reduces the single-top-quark and $t\bar{t}$ backgrounds. Events are further categorized into VBF
(with a focus on the VBF production of the Higgs boson) and non-VBF
categories. The VBF selection is
based on the kinematics of the two jets with the highest $\pt$, where $\mathrm{j}_1$ and $\mathrm{j}_2$ denote the leading and subleading jet in \pt, respectively.
The variables {\mjj} and {\detajj} stand for the invariant mass and
$\eta$ separation of these two jets.
The non-VBF category contains events failing the VBF selection.
In the dilepton channel, additional selection criteria are applied to further reject background events in this category. These criteria are also listed in Table~\ref{tab:event_selection}, where {\mt} stands for the transverse
mass\footnote{The transverse mass of two objects is defined as
$\mt = \sqrt{2 {\pt}_1 {\pt}_2 (1-\cos\Delta\phi)}$, where ${\pt}_i$
are the individual transverse momenta and $\Delta\phi$ is the angle
between the two objects in the azimuthal plane.} of the two objects
listed in parentheses, and $\pt^{\tau}$ represents the magnitude of
the vector sum of $\pt^{\ell_2}$ and {\MET}. The requirement on
$\pt^{\tau}/\pt^{\ell_1}$ reduces the background arising
from jets misidentified as leptons.
The VBF and non-VBF categories in each of the {\leplep} and {\lephad} channels give rise to four signal regions in each search.

The analysis exploits
BDT algorithms to enhance the signal separation from the background in
the individual searches, channels and categories.
The components of the four-momenta of the analysis objects as well as derived event
variables (e.g.\ invariant masses and angular separations) are the input variables of the BDT discriminant.
Correlations between these input variables have been carefully checked, highly
correlated variables have been removed and the remaining ones are ranked according to their discrimination power~\cite{TMVA,XGBoost}. The list of variables is then optimized, removing the lowest-ranked variables with marginal contribution to the sensitivity.
The final list of variables is presented in Table~\ref{tab:BDT_input_vars} for each channel and category.
The invariant mass of the Higgs boson reconstructed under the  $H\to\ell\tau$ decay hypothesis exhibits the highest signal-to-background
separation power and it helps to distinguish LFV signal from $H\to\tau\tau$ and $H\to WW$ backgrounds.
For the $\leplep$ channel the invariant mass is reconstructed with the MMC algorithm~\cite{MMCpaper} and is denoted by $m_{\text{MMC}}$; for the $\lephad$ channel it is reconstructed with the collinear approximation~\cite{MMCpaper} and is denoted by $m_{\text{coll}}$.
The analysis CRs are used to validate the level of agreement between data and simulated distributions of the BDT score and input variables, as well as their correlations.
 
\begin{table}[hbt]
\caption{BDT input variables used in the analysis. For each channel and
category, used input variables are marked with HR (indicating the
five variables with the highest rank) or a bullet. Analogous
variables between the two channels are listed on the same line.
}
\label{tab:BDT_input_vars}
\centering
\small{
\setlength{\extrarowheight}{1pt}
\begin{tabular}{ccc|ccc}
\toprule
\multicolumn{3}{c|}{\leplep} & \multicolumn{3}{c}{\lephad} \\
Variable & VBF & non-VBF & Variable & VBF & non-VBF \\
\midrule
$m_{\text{MMC}}$           & HR & HR & $m_{\text{coll}}$
& HR & HR \\
$\pt^{\ell_1}$             & $\bullet$ & $\bullet$  & $\pt^{\ell}$
& $\bullet$ & HR \\
$\pt^{\ell_2}$             & HR & HR & $\pt^{\tauhadvis}$
& $\bullet$ & HR \\
$\Delta R(\ell_1,\ell_2)$ & HR & $\bullet$ & $\Delta R(\ell,\tauhadvis)$
& $\bullet$ & $\bullet$ \\
$\mt(\ell_1,\MET)$        & $\bullet$  & HR & $\mt(\ell,\MET)$     & HR & $\bullet$ \\
$\mt(\ell_2,\MET)$        & HR &  $\bullet$ & $\mt(\tauhadvis,\MET)$
& HR   & HR \\
$\Delta\phi(\ell_1,\MET)$ &  $\bullet$  &  $\bullet$ & $\Delta\phi(\ell,\MET)$
&  HR  & $\bullet$ \\
$\Delta\phi(\ell_2,\MET)$ &  & HR & $\Delta\phi(\tauhadvis,\MET)$
& $\bullet$   &  \\
\mjj                      &  $\bullet$  &         & \mjj & $\bullet$ & \\
\detajj                   & HR &        & \detajj & $\bullet$ & \\
$\pt^{\tau}/\pt^{\ell_1}$   &    & HR & $\sum\limits_{i=\ell,\tauhadvis}\!\!\!\!\!\!\!\!\cos\Delta\phi(i,\MET)$
& $\bullet$ & $\bullet$ \\
& & & \MET & HR & $\bullet$ \\
& & & \mvis & & HR \\
& & & $\Delta\eta(\ell,\tauhadvis)$ &  & $\bullet$ \\
& & & $\eta^{\ell}$ &  & $\bullet$ \\
& & & $\eta^{\tauhadvis}$ & & $\bullet$ \\
& & & $\phi^{\ell}$ & & $\bullet$ \\
& & & $\phi^{\tauhadvis}$ & & $\bullet$ \\
& & & $\phi(\MET)$  & & $\bullet$ \\
\bottomrule
\end{tabular}
}
\end{table}
 
\FloatBarrier
\section{Background modelling}
\label{sec:background}
The most significant backgrounds in the search are from events with
$Z\to\tau\tau$ decays or with (single or pair-produced) top quarks, especially in the {\leplep} channel,
as well as from events with misidentified objects, which are estimated using data-driven (d.d.) techniques. The relative contribution from misidentified objects to the total background yield
is 5--25\% in the {\leplep} channel and 25--45\% in the  {\lephad} channel, depending on the search and the analysis category.
The shapes of distributions from the $Z\to\tau\tau$ and top-quark (single-top-quark and $t\bar{t}$) processes are modelled by simulation in both the {\leplep} and {\lephad} decay channels.
In the {\leplep} channel, the relative contributions of $Z\to\tau\tau$ and  top-quark production processes are 20--35\% and 20--55\%, respectively; the top-quark background dominates in the VBF category. In the {\lephad} channel, the top-quark background fraction is 1--10\%, while the $Z\to\tau\tau$ process contributes to 45--55\% of the total background. The individual contributions are listed in Tables~\ref{tab:tab_yield_etau} and~\ref{tab:tab_yield_mtau}.
Smaller background components are also modelled by simulation and are grouped together: $Z\to\mu\mu$, diboson production, $H\rightarrow\tau\tau$ and $H\rightarrow WW$.
 
Good modelling of the background is demonstrated in
Figure~\ref{fig:prefit_dist} for a selection of important BDT input variables.
Details of the background estimation techniques are given below.
 
\begin{figure}[htb]
\centering
\includegraphics[width=0.32\textwidth]{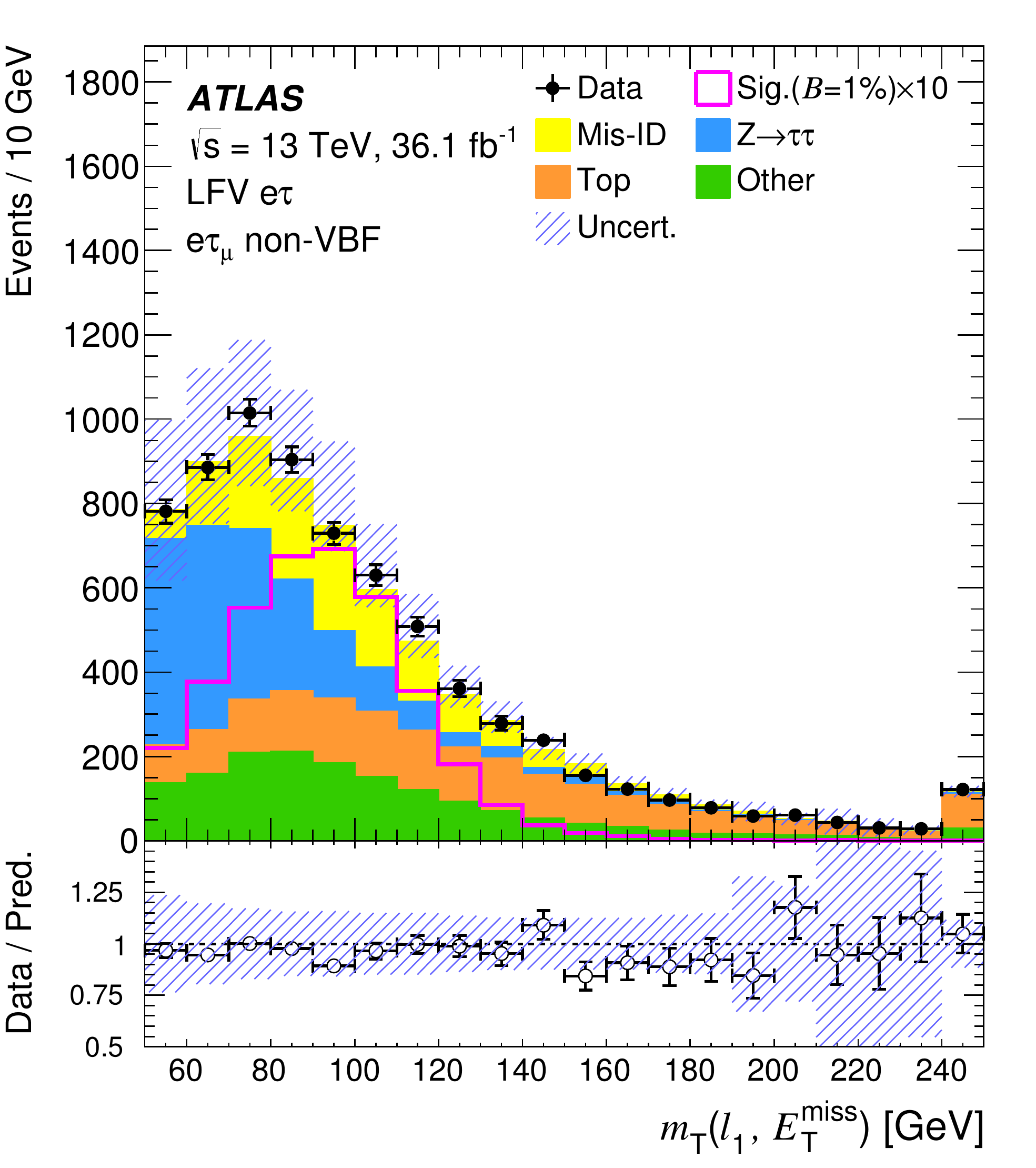}
\includegraphics[width=0.32\textwidth]{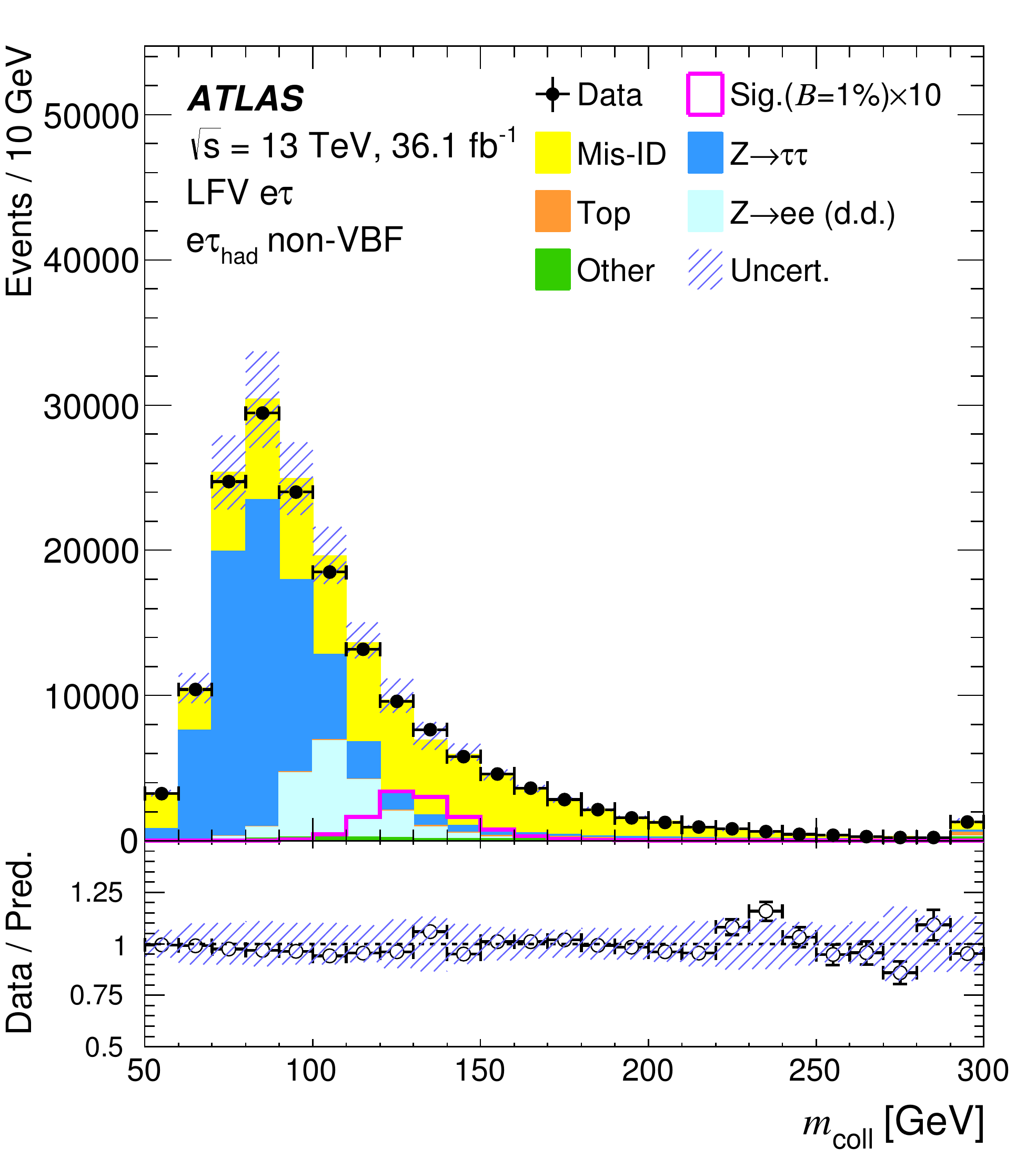}
\includegraphics[width=0.32\textwidth]{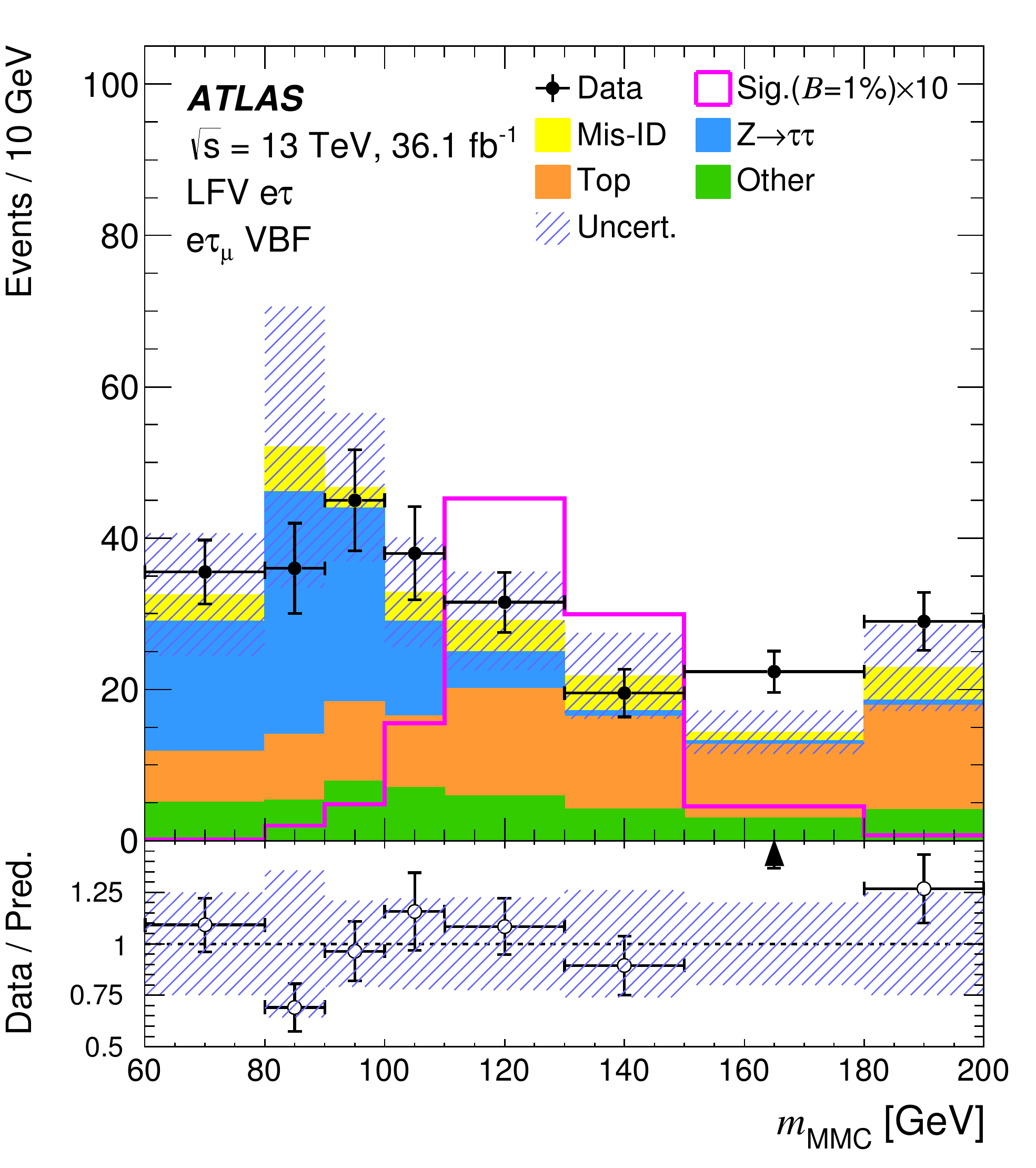} \\
\includegraphics[width=0.32\textwidth]{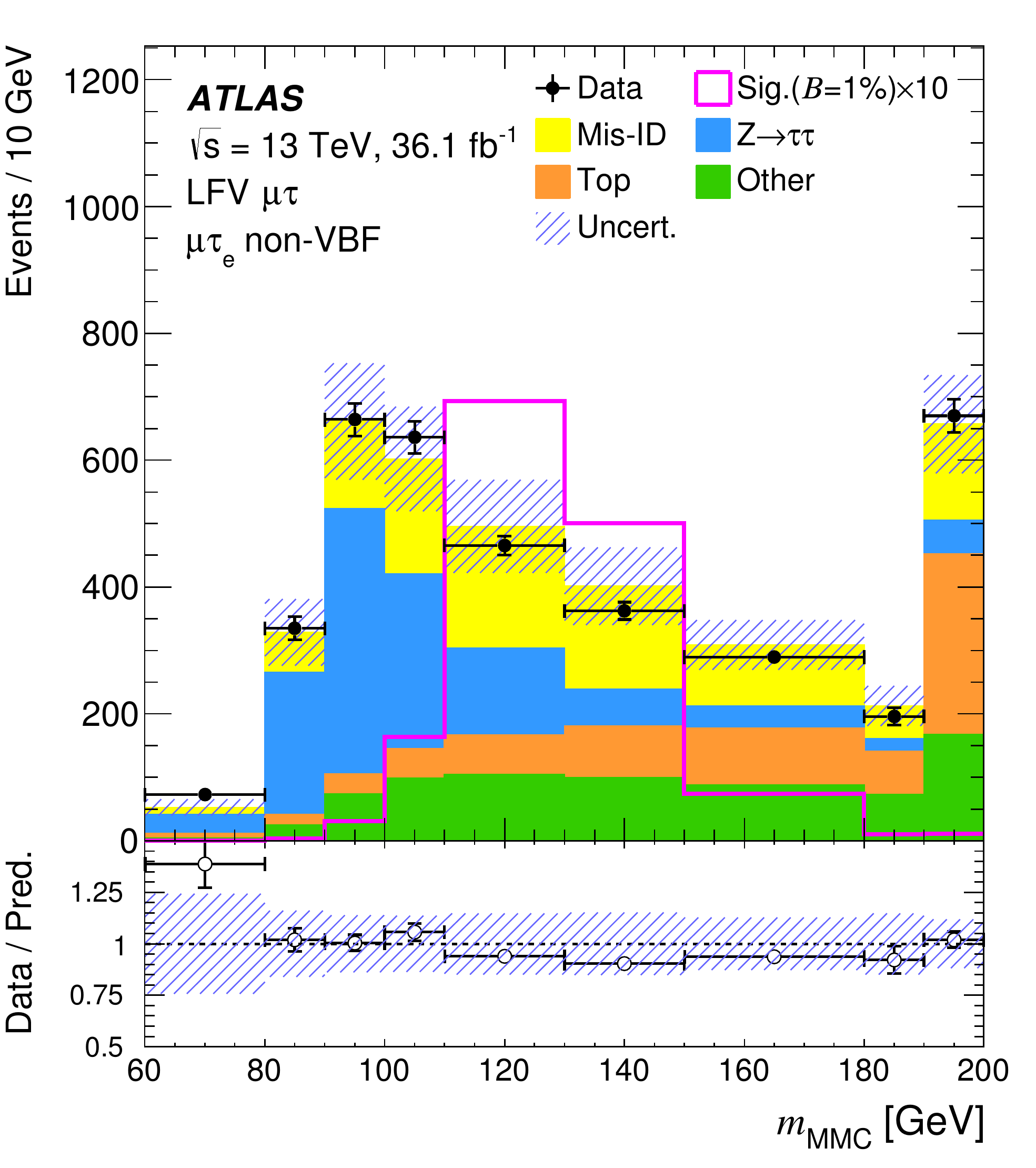}
\includegraphics[width=0.32\textwidth]{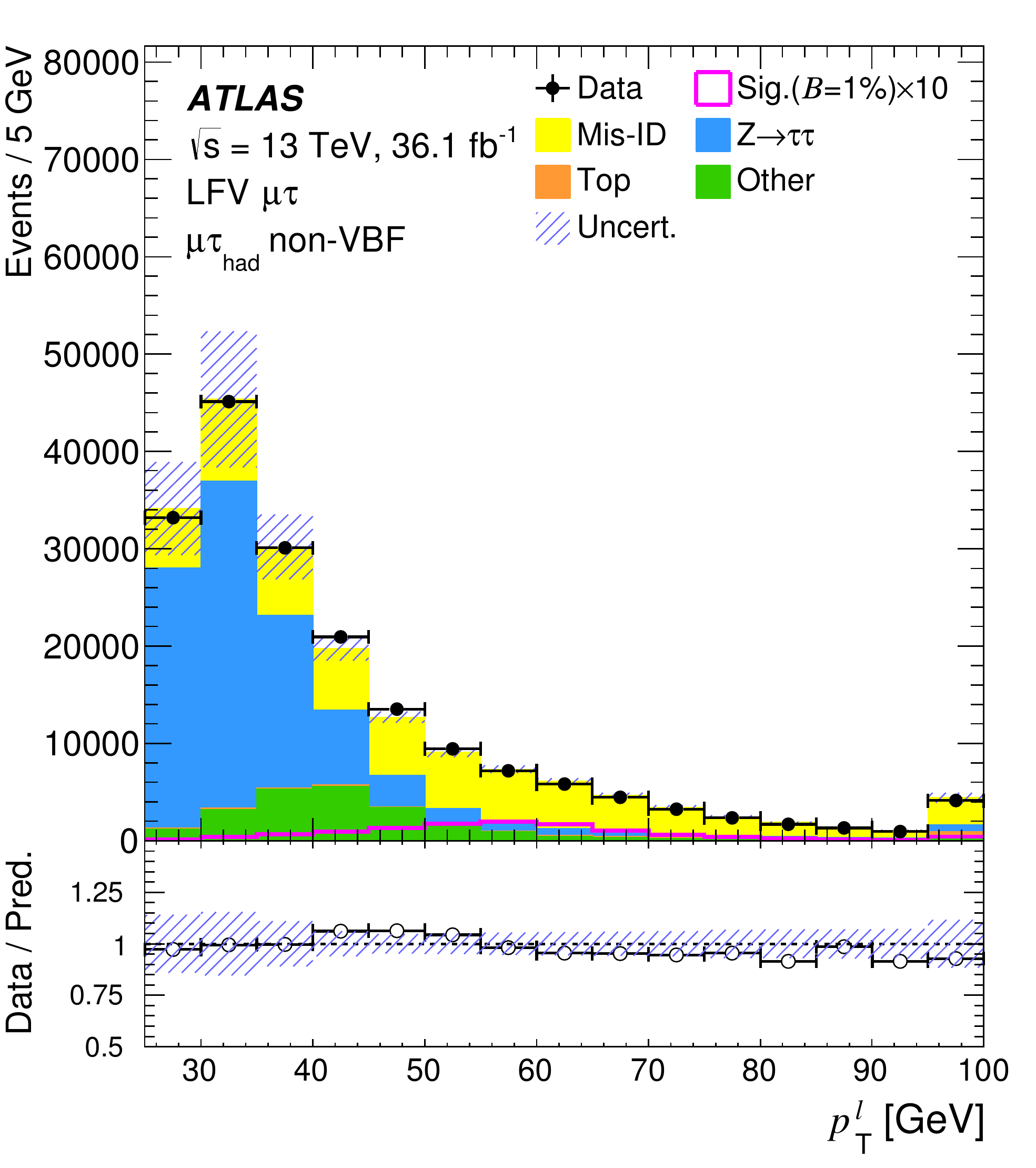}
\includegraphics[width=0.32\textwidth]{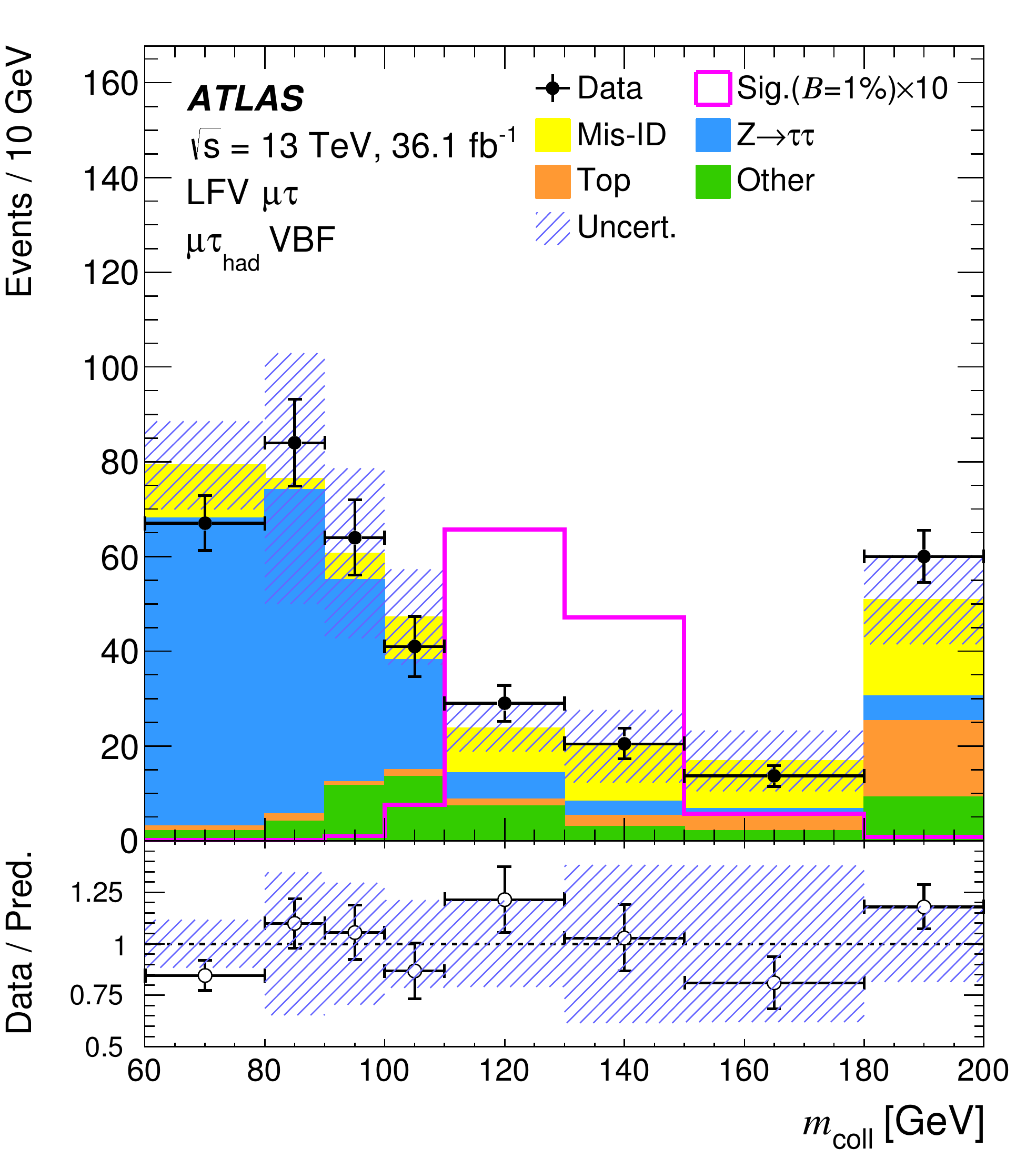}
\caption{Distributions of representative kinematic
quantities for different searches, channels and categories, before the fit
as described in Section~\ref{sec:stat_ana} is applied.
Top row: transverse mass $\mt(\ell_1,\MET)$ ($e\tau_\mu$ non-VBF),
collinear mass $m_{\text{coll}}$ ($e\tau_{\text{had}}$ non-VBF) and
$m_{\text{MMC}}$ ($e\tau_\mu$ VBF).
Bottom row: $m_{\text{MMC}}$ ($\mu\tau_e$ non-VBF), muon {\pt}
($\mu\tau_{\text{had}}$ non-VBF) and $m_{\text{coll}}$
($\mu\tau_{\text{had}}$ VBF).
Entries with values that would exceed the $x$-axis range are included
in the last bin of each distribution. The size of the combined statistical,
experimental and theoretical uncertainties in the background is indicated
by the hatched bands.
The $H\to e\tau$ ($H\to\mu\tau$) signal overlaid in top (bottom)
plots assumes $\mathcal{B}(H\to\ell\tau)=1\%$ and is
enhanced by a factor $10$.
In the data/background prediction ratio plots, points outside the
displayed $y$-axis range are shown by arrows.
}
\label{fig:prefit_dist}
\end{figure}
 
\subsection{{\leplep} channel}
Two sets of CRs, as defined in Table~\ref{tab:event_selection}, are used to constrain the normalization of $Z\to\tau\tau$ and top-quark background components.
These CRs inherit their definitions from the corresponding analysis
category but invert one requirement
to ensure orthogonality with the nominal selection. The normalization factors are determined during the statistical
analysis by fitting the event yields in all signal and control regions simultaneously.
For each search, separate $Z\to\tau\tau$ normalization factors are used for the VBF and non-VBF categories.
In the case of the top-quark background, in which leading jets are produced at a lower order of the perturbative expansion of the scattering process, a combined normalization factor across the two categories is used in the {\leplep} channel.
 
Top-quark CRs are almost exclusively composed of top-quark backgrounds: the purity is 95\% across both searches and categories, with $t\bar{t}$ process accounting for more than 90\% of the top-quark backgrounds.
The $Z\to\tau\tau$ CRs achieved a purity of $\sim$80\% in the non-VBF categories, while a lower purity of $\sim$60\% is observed in the VBF categories.
The contributions of all other background components
are normalized to their SM predictions when the likelihood fit (Section~\ref{sec:stat_ana}) is applied.
 
The shape and normalization of diboson and $Z\to\mu\mu$ background distributions
are validated with data in dedicated regions where their contributions are
enhanced. The latter process only contributes sizeably in the
$\mu\tau_{e}$ channel,
where it represents up to 10\% of the total background.
 
Another source of background comes from $W+$jets, top-quark and multi-jet events,
where jets are misidentified as leptons.
This background is estimated directly from OS data events where an inverted isolation requirement is imposed on the subleading lepton~\cite{htautau}. Normalization factors are applied to correct for the inverted isolation requirement. The normalization factors are derived in a dedicated region where the leptons are required to
have same-sign (SS) charges.
Additional corrections are made by reweighting the MC distributions of $\Delta\phi(\ell_1,\MET)$ and $\Delta\phi(\ell_2,\MET)$ to data in the SS region, which improves the modelling of azimuthal angles between leptons and the {\MET} direction as well as the modelling of $\mt(\ell_2,\MET)$.
A similar improvement is observed in the nominal OS region.
In most of the cases, the misidentified jet mimics the lepton of lower $\pt$, $\ell_{2}$, while the fraction of events where both leptons are misidentified varies between 2\% to 8\% across categories.
The systematic uncertainties of the estimation of the misidentified lepton background include contributions from closure tests in SS and OS regions enriched with misidentified leptons, from the corrections made to the $\Delta\phi$ distributions, and from the composition of the misidentified lepton background.
 
\subsection{{\lephad} channel}
The main background contributions come from the $Z\to\tau\tau$ process and events where either a jet or an electron is misidentified as {\tauhadvis}.
The shape of the $Z\to\tau\tau$ background distribution is modelled by simulation, and the
corresponding normalization factors are determined from the
simultaneous fit of the event yields in all signal and control
regions. The $Z\to\tau\tau$ normalization factors are fully correlated
with those of the {\leplep} channel, in each VBF and non-VBF
category. Top-quark production represents less than {1\%} of the total background in the {\lephad} channel and is determined by simulation, including its normalization, which is kept fixed in the fit.

The main contributions to jets misidentified as {\tauhadvis} come from multi-jet events and $W$-boson production in association with jets, and a fake-factor method is used to estimate the contribution of each component separately. A fake factor is defined as the ratio of the number of events where the highest-{\pt} jet is identified as a tight {\tauhadvis} candidate to the number of events where the highest-{\pt} jet fails to satisfy this $\tau$-ID criterion but satisfies a looser criterion. The procedure, including systematic uncertainties, is described in Ref.~\cite{htautau}. Since a different $\tau$-ID working point is considered in this analysis, fake factors are re-derived as a function of {\pt} and track multiplicity of the {\tauhadvis} candidate.
 
Electrons misidentified as {\tauhadvis}, denoted by ``$Z\to ee$ (d.d.)''
in the following figures and tables, represent another background component in the $e\thad$ channel, with a contribution about five times smaller than that of jets misidentified as {\tauhadvis}.
While the rate of electrons  misidentified as 3-prong {\tauhadvis} makes a negligible contribution and is modelled by simulation, the rate of electrons misidentified as 1-prong {\tauhadvis} is
determined with a fake-factor method. This time, the fake factor is defined
as the ratio of the number of events with tight $\tau$-ID to the number of
events with anti-identified {\tauhadvis} (such a candidate satisfies all criteria
but the requirement on the high electron ID score is inverted).
These fake factors are derived in a dedicated $Z\to e e$ enriched region defined by $|\mvis - m_Z| < 5\,\GeV$, $\mt(\ell,\MET) < 40\,\GeV$, and
$\mt(\tauhadvis,\MET) < 60\,\GeV$, where the {\tauhadvis} candidate satisfies
the medium $\tau$-ID (corresponding to an efficiency of 55\% and 40\% for 1-prong and 3-prong candidates, respectively) but not the tight $\tau$-ID criterion to avoid overlap with the {\lephad} signal region. These fake factors are applied to signal-like events with the anti-identified {\tauhadvis} to determine the background contribution in the categories of the analysis.
The systematic uncertainties include the statistical uncertainty of
the fake factors and account for looser $\tau$-ID in the $Z\to e e$
enriched region as well as for the subtraction of the not
misidentified components in this region.
 
\FloatBarrier
\section{Systematic uncertainties}
The systematic uncertainties affect the normalization of signal and background, and/or the shape of their corresponding final discriminant distributions.
Each source of systematic uncertainty is considered to be uncorrelated with the other sources. The effect of each systematic uncertainty is fully considered in each category, including control regions. Correlations of each systematic uncertainty are maintained across processes, channels, categories and regions.
The size of the systematic uncertainties and their impact on the fitted branching ratio
are discussed in Section~\ref{sec:stat_ana}. The main sources of systematic uncertainties are related to the estimation of the backgrounds originating from mis-identified leptons/jets and to the jet energy scale uncertainties.
 
Experimental uncertainties include those originating from the reconstruction, identification, tagging and triggering efficiencies of all physics objects as well as their momentum scale and resolution.
These include effects from leptons~\cite{PERF-2015-10,PERF-2017-03,PERF-2017-01}, {\tauhadvis}~\cite{ATLAS-CONF-2017-029}, jets~\cite{PERF-2016-04,ATLAS-CONF-2014-018,PERF-2016-06} and $\MET$~\cite{PERF-2016-07}.
Uncertainties affecting the kinematics of the physics objects are propagated to the BDT input variables. The corresponding shape and normalization variations of the BDT discriminant are considered in the statistical analysis.
Uncertainties of the luminosity measurement~\cite{ATLAS-CONF-2019-021}, pile-up modelling and uncertainties specific to mis-identified background estimation techniques mentioned in Section~\ref{sec:background} are included.
 
The procedures to estimate the uncertainty of the Higgs boson production cross-sections follow the recommendations of the LHC Higgs Cross-Section Working Group \cite{theo_sys_1}.
Theoretical uncertainties affecting the ggF signal originate from nine sources~\cite{LHCXS}.
Two sources account for yield uncertainties, which are evaluated by an overall variation of all relevant scales and are correlated across all bins of the BDT discriminant distribution~\cite{theo_sys_3}. Another two sources account
for migration uncertainties of zero to one jet and one to at least two
jets in the event~\cite{theo_sys_3,theo_sys_4,theo_sys_5}, two for
Higgs boson $\pt$ shape uncertainties, one for the treatment of the
top-quark mass in the loop corrections, and two for the acceptance
uncertainties of ggF production in the VBF phase space from selecting
exactly two and at least three jets, respectively
\cite{theo_sys_7,theo_sys_8}. For VBF and $WH$, $ZH$ production
cross-sections, the uncertainties due to missing higher-order QCD
corrections are estimated by varying the factorization and
renormalization scales up and down by factors of two around the nominal scale.
For all signal samples, PDF uncertainties are estimated using 30 eigenvector variations and two $\alpha_{\mathrm{s}}$ variations using the default PDF set PDF4LHC15 \cite{Butterworth:2015oua}. Uncertainties related to the simulation of the underlying event, hadronization and parton shower are estimated by comparing the acceptances when using Pythia~8.212~\cite{Sjostrand:2014zea} or Herwig~7.0.3~\cite{herwig_1,herwig_2}.
 
The sources of modelling uncertainties considered for the
$Z\to\tau\tau$ process are the same as in Ref.~\cite{htautau} and
their effect on the event migrations between categories and on the
shape of the BDT discriminant are considered, since the overall
normalizations are determined from data in the statistical
analysis. These systematic uncertainties include variations of PDF
sets, factorization and renormalization scales, CKKW matching
\cite{CKKW}, resummation scale and parton shower modelling. The other
background processes are either normalized using data (processes with
top-quarks and mis-identified leptons and {\tauhadvis} candidates) or their
cross-section uncertainties have negligible impact and therefore are not included. The shape uncertainties of these backgrounds originate from experimental uncertainties only.
\section{Statistical analysis}
\label{sec:stat_ana}
The searches for $H\to e\tau$ and $H\to\mu\tau$ are treated independently.
For each search, the analysis exploits the four signal regions and the two
control regions specified in Table~\ref{tab:event_selection}. The BDT score distributions of all signal regions are analysed to test the presence of a signal, simultaneously with the event yields in control regions, which are included to constrain the normalizations of the major backgrounds estimated from simulation.  The statistical analysis uses a binned likelihood function $\mathcal{L}(\mu,\theta)$, constructed as a product of Poisson probability terms over all bins considered in the search. This function depends on the parameter $\mu$, defined as the branching ratio $\mathcal{B}(H\to\ell\tau)$, and a set of nuisance parameters $\theta$ that encode the effect of systematic uncertainties in the signal and background expectations. All nuisance parameters are implemented in the likelihood function as Gaussian or log-normal constraints.
The normalization factors of the single-top-quark and $t\bar{t}$
backgrounds in the {\leplep} channel and of the $Z\to\tau\tau$ background component are unconstrained parameters of the fit.
Estimates of the parameters of interest are calculated with the
profile-likelihood-ratio test statistic $\tilde{q}_{\mu}$~\cite{Cowan:2010js},
and the upper limits on the branching ratios are derived by using
$\tilde{q}_{\mu}$ and the CL$_{\mathrm{S}}$ method~\cite{Read:2002hq}.
 
The discriminant distributions after the fit in each
channel are shown in Figures~\ref{fig:lfv_mva_postfit_bdt_etau}
and~\ref{fig:lfv_mva_postfit_bdt_mutau}. Good agreement between data and
the background expectation is observed.
The event yields after the background-only fit are summarized in
Tables~\ref{tab:tab_yield_etau} and~\ref{tab:tab_yield_mtau}.
In the non-VBF category, the yields in the {\lephad} channel are larger than
in the {\leplep} channel due
to the looser selection criteria defined for the former channel
(Section~\ref{sec:event_selection}).
Table~\ref{tab:NPimpactMVA} shows a summary of the
uncertainties of $\mathcal{B}(H\to\ell\tau)$.
The uncertainties associated with misidentified leptons and {\tauhadvis} candidated and
those related to the jet energy scale and resolution exhibit
the highest impact on the best-fit branching ratios in both searches.
The combined impact from all systematic uncertainties and the data statistics ranges from $0.17\%$ to $0.19\%$.
 
\begin{table}[htb]
\caption{Event yields and predictions as determined by the background-only fit in different signal regions of the $H\to e\tau$ analysis. Uncertainties include both the statistical and systematic contributions.
``Other'' contains diboson, $Z\to\ell\ell$, $H\to\tau\tau$ and $H\to WW$ background processes.
For the $e\thad$ channel the ``$Z\to ee$ (d.d.)'' component corresponds to electrons misidentified as {\tauhadvis}. This contribution is summed with ``Other'' since there are few events in the VBF category.
The uncertainty of the total background includes all correlations
between channels. The normalizations of top-quark ({\leplep} channel
only) and $Z\to\tau\tau$ background components are determined by the
fit, while the expected signal event yields are given for
$\mathcal{B}(H\to e\tau) = 1\%$.}
\centering
\setlength{\extrarowheight}{0pt}
\begin{tabular}{lD{+}{\,\pm\,}{5,3}D{+}{\,\pm\,}{5,3}D{+}{\,\pm\,}{5,3}D{+}{\,\pm\,}{5,3}}
\toprule
& \multicolumn{1}{c}{$e\tau_\mu$ non-VBF}
& \multicolumn{1}{c}{$e\tau_\mu$ VBF}
& \multicolumn{1}{c}{$e\tau_{\text{had}}$ non-VBF}
& \multicolumn{1}{c}{$e\tau_{\text{had}}$ VBF} \\ \midrule
Signal & 379 + 31 & 19.8 + 2.7 & 1180 + 110 & 25 + 4  \\ \midrule
$Z\to\tau\tau$ & 2470 + 230 & 221 + 34 & 73\,800 + 1900 & 290 + 40 \\
Top-quark & 1640 + 140 & 490 + 40 & 1580 + 190 & 56 + 12 \\
Mis-identified & 1330 + 250 & 73 + 33 & 74\,400 + 1600 & 140 + 50 \\
$Z\to ee$ (d.d.) & \multicolumn{1}{c}{\multirow{2}{*}{}}
& \multicolumn{1}{c}{\multirow{2}{*}{$\phantom{1}$}}
& 15\,900 + 1800 & \multicolumn{1}{c}{\multirow{2}{*}{$\phantom{12}82\pm 13$}} \\
Other & 1700 + 80 & 220 + 15  & 2960 + 200 &  \\ \midrule
Total background & 7130 + 100 & 1003 + 33 & 168\,700 + 1000 & 570 + 40 \\
Data & \multicolumn{1}{c}{$7128\phantom{\,\pm\,10}$} & \multicolumn{1}{c}{$992\phantom{\,\pm\,1}$}
& \multicolumn{1}{c}{$168\,883\phantom{\,\pm\,1000}$} & \multicolumn{1}{c}{$572\phantom{\,\pm\,1}$} \\
\bottomrule
\end{tabular}
\label{tab:tab_yield_etau}
\end{table}
 
\begin{table}[htb]
\caption{Event yields and predictions as determined by the
background-only fit in different signal regions of the $H\to\mu\tau$
analysis. Uncertainties include both the statistical and systematic
contributions.
``Other'' contains diboson, $Z\to\ell\ell$, $H\to\tau\tau$ and $H\to WW$ background processes.The uncertainty of the total background includes all correlations between channels.
The normalizations of top-quark ({\leplep} channel only) and
$Z\to\tau\tau$ background components are determined by the fit, while
the expected signal event yields are given for
$\mathcal{B}(H\to\mu\tau) = 1\%$.}
\centering
\begin{tabular}{lD{+}{\,\pm\,}{5,3}D{+}{\,\pm\,}{5,3}D{+}{\,\pm\,}{5,3}D{+}{\,\pm\,}{5,3}}
\toprule
& \multicolumn{1}{c}{$\mu\tau_e$ non-VBF}
& \multicolumn{1}{c}{$\mu\tau_e$ VBF}
& \multicolumn{1}{c}{$\mu\tau_{\text{had}}$ non-VBF}
& \multicolumn{1}{c}{$\mu\tau_{\text{had}}$ VBF} \\ \midrule
Signal & 287 + 23 & 14.6 + 1.9 & 1200 + 120 & 25 + 5 \\ \midrule
$Z\to\tau\tau$ & 1860 + 130 & 144 + 26 & 96\,100 + 2000 & 274 + 33 \\
Top quark & 1260 + 130 & 390 + 34 & 1620 + 210 & 51 + 10 \\
Misidentified & 1340 + 210 & 41 + 21 & 63\,900 + 1600 & 149 + 33 \\
Other & 1180 + 140 & 168 + 18 & 23\,000 + 1000 & 104 + 15 \\ \midrule
Total background & 5640 + 100 & 743 + 29 & 184\,500 + 1200 & 580 + 30 \\
Data & \multicolumn{1}{c}{$5664\phantom{\,\pm\,10}$} & \multicolumn{1}{c}{$723\phantom{\,\pm\,1}$}
& \multicolumn{1}{c}{$184\,508\phantom{\,\pm\,1000}$} & \multicolumn{1}{c}{$583\phantom{\,\pm\,1}$} \\ \bottomrule
\end{tabular}
\label{tab:tab_yield_mtau}
\end{table}
 
\begin{figure}[htb]
\centering
\includegraphics[width=0.40\textwidth]{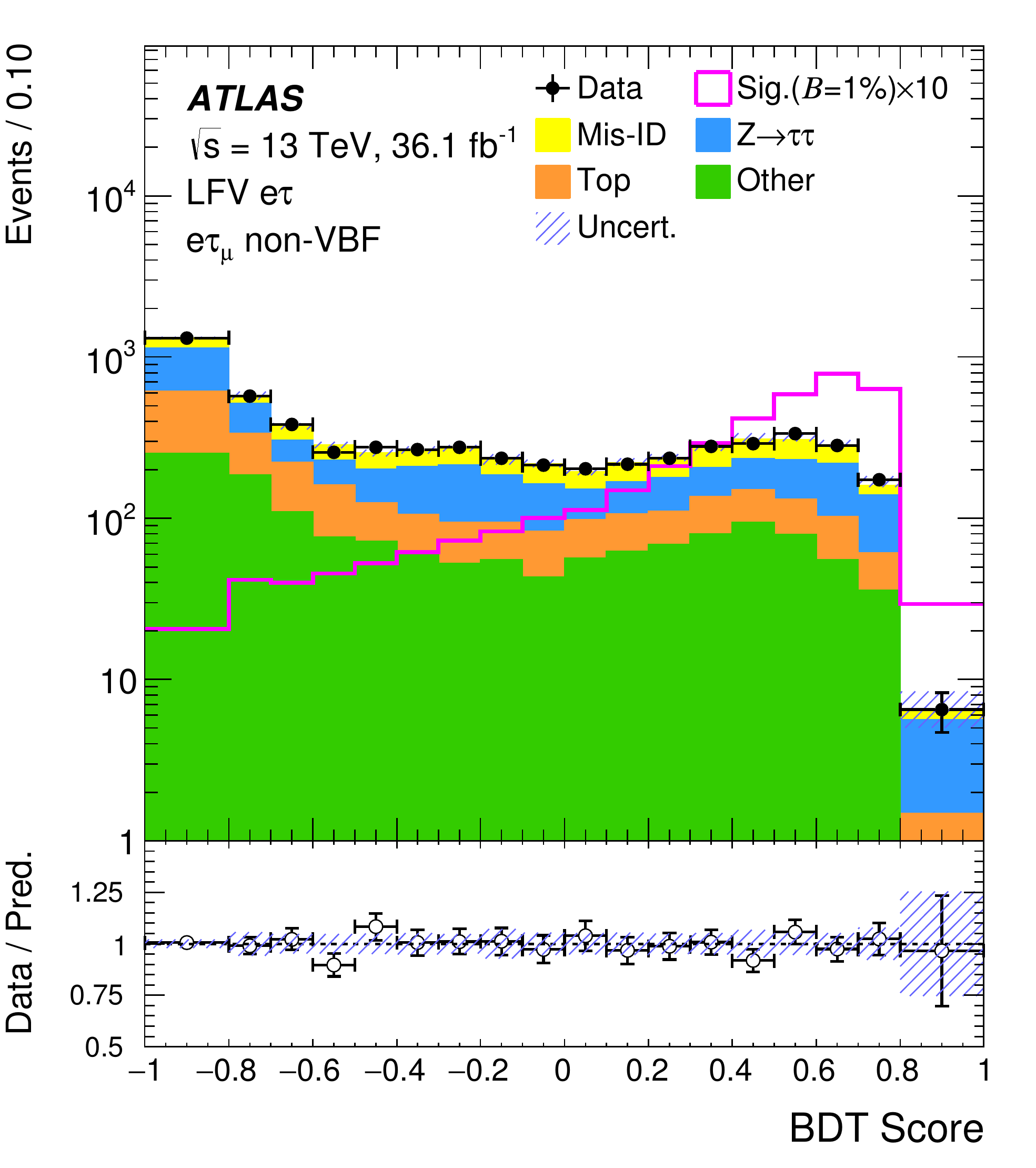}
\includegraphics[width=0.40\textwidth]{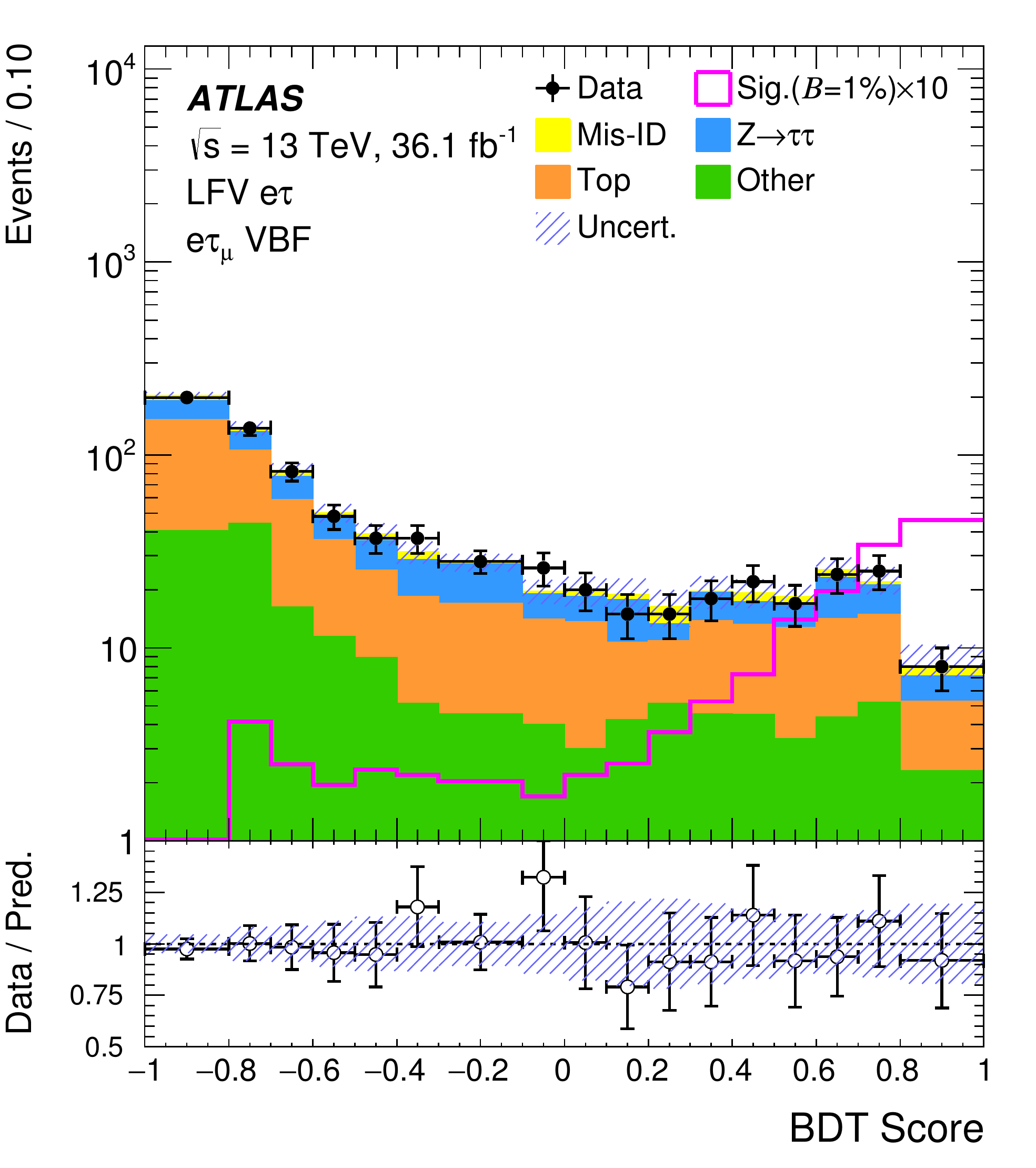}\\
\includegraphics[width=0.40\textwidth]{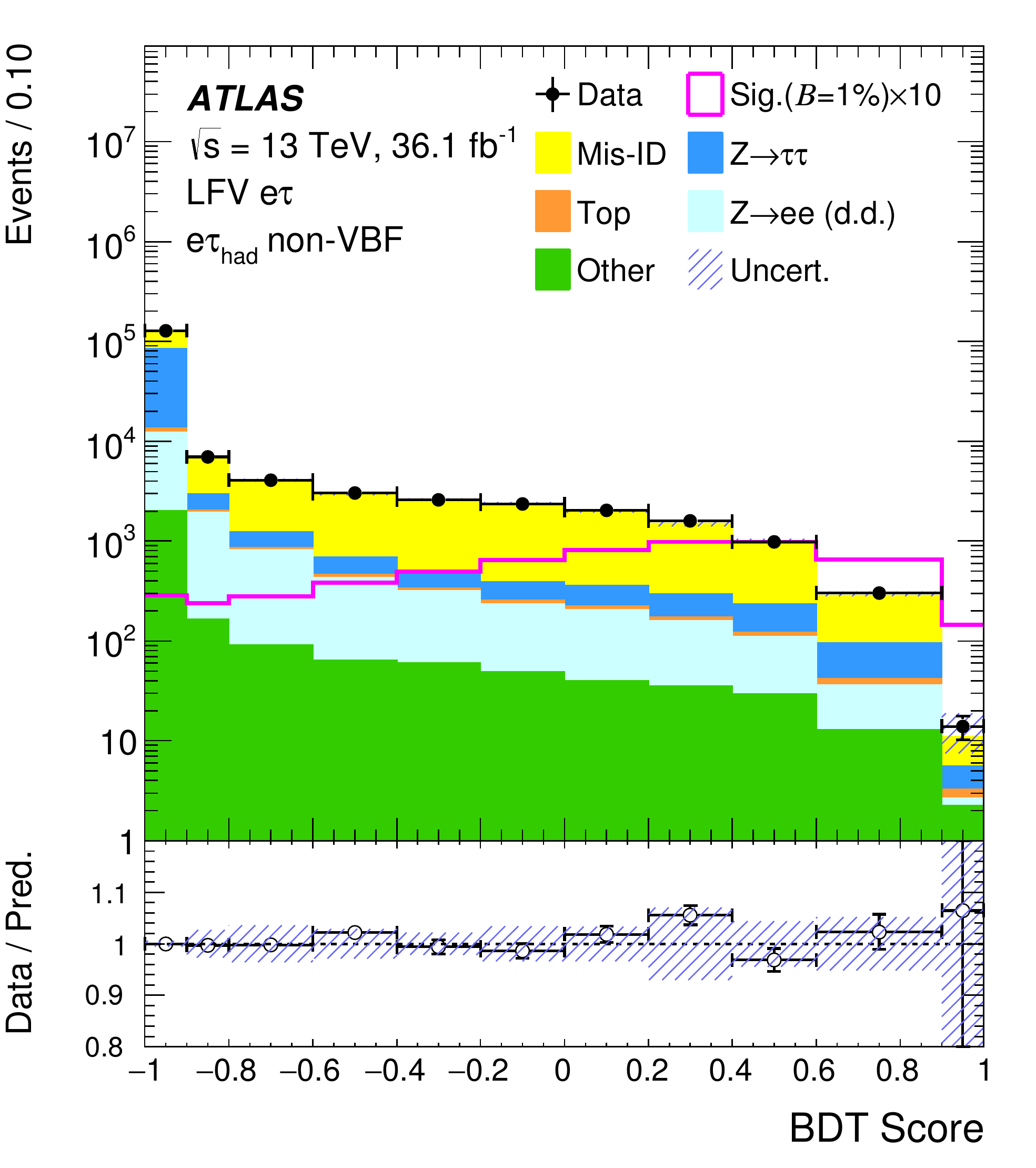}
\includegraphics[width=0.40\textwidth]{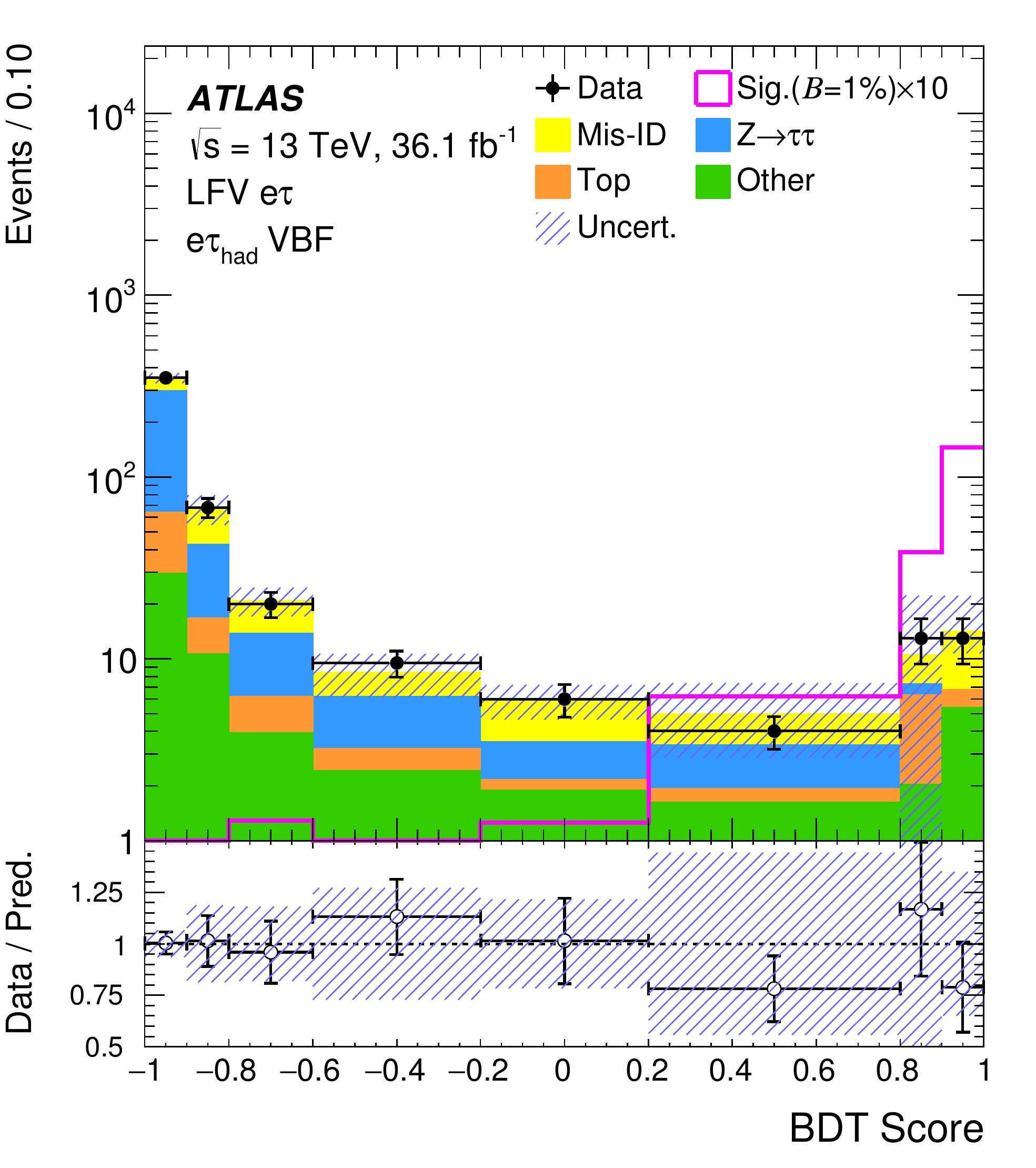}
\caption{Distributions of the BDT score after the background+signal fit in
each signal region of the $e\tau$ search, with the LFV signal
overlaid, normalized with $\mathcal{B}(H\to e\tau) = 1\%$
and enhanced by a factor $10$ for visibility.
The top and bottom plots display $e\tau_{\mu}$ and $e\tauhad$ BDT scores
respectively, the left (right) column corresponds to the non-VBF (VBF)
category. The size of the combined statistical,
experimental and theoretical uncertainties of the background is indicated
by the hatched bands.
The binning is shown as in the statistical analysis.
}
\label{fig:lfv_mva_postfit_bdt_etau}
\end{figure}

\begin{figure}[htb]
\centering
\includegraphics[width=0.40\textwidth]{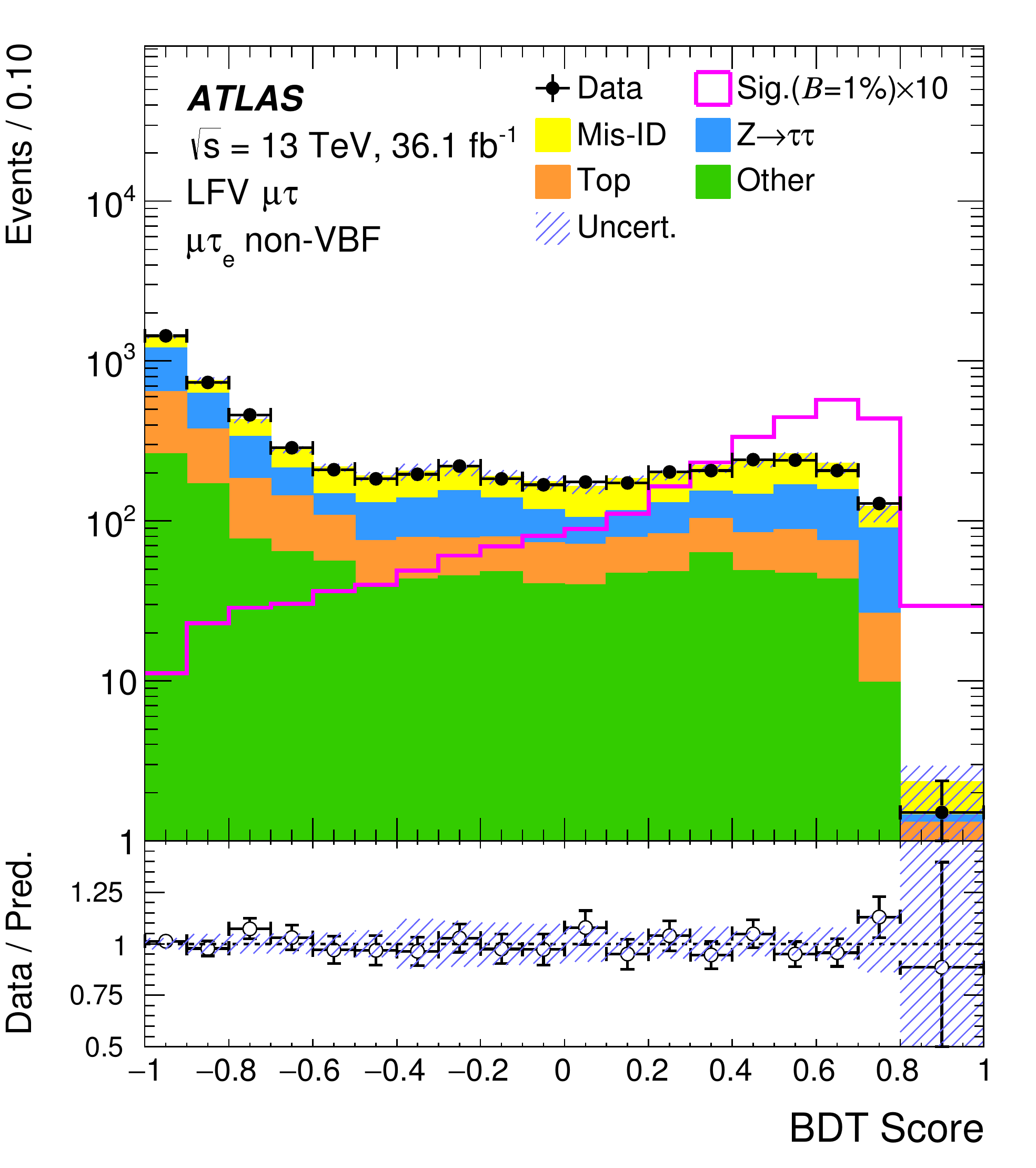}
\includegraphics[width=0.40\textwidth]{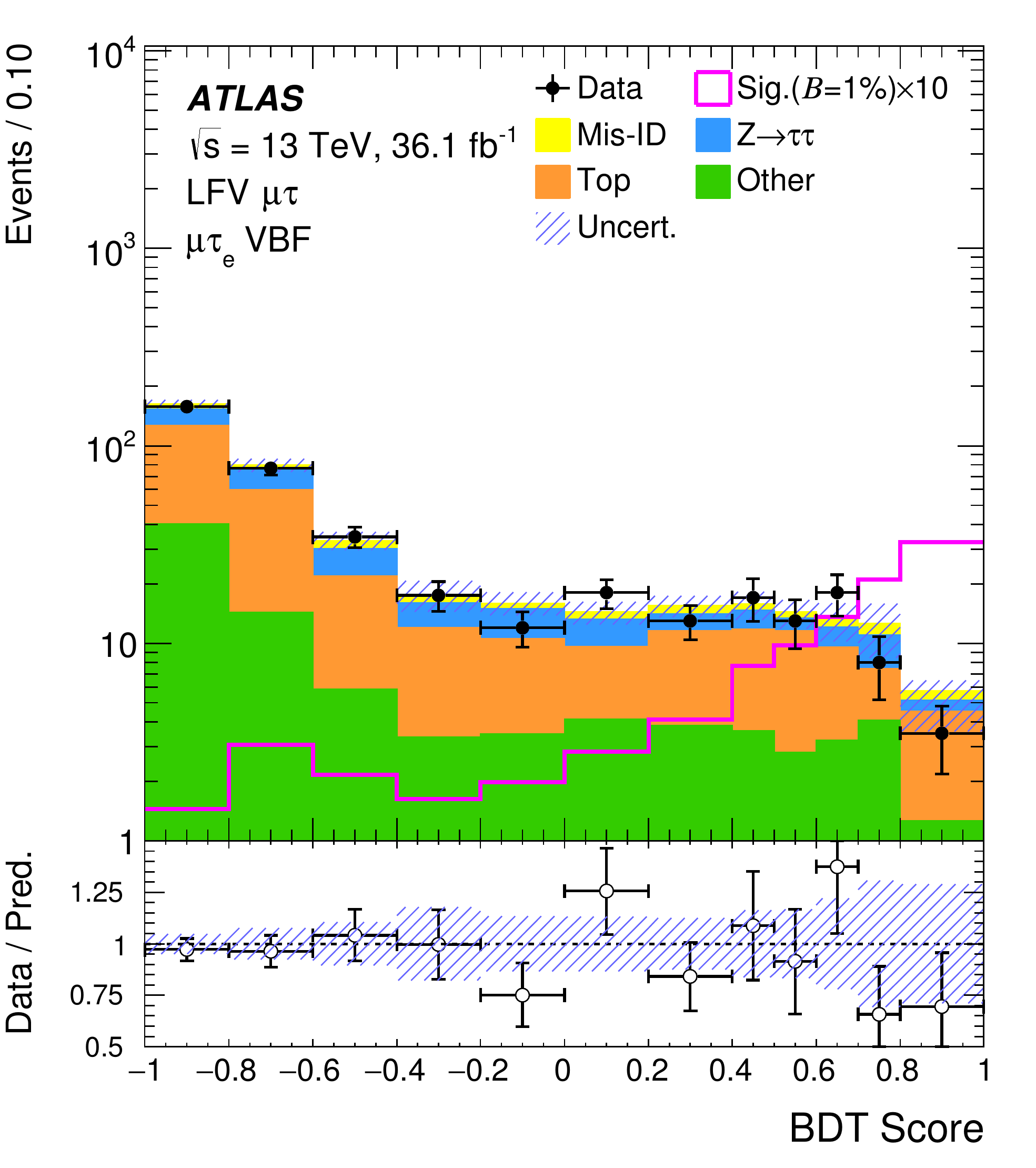}\\
\includegraphics[width=0.40\textwidth]{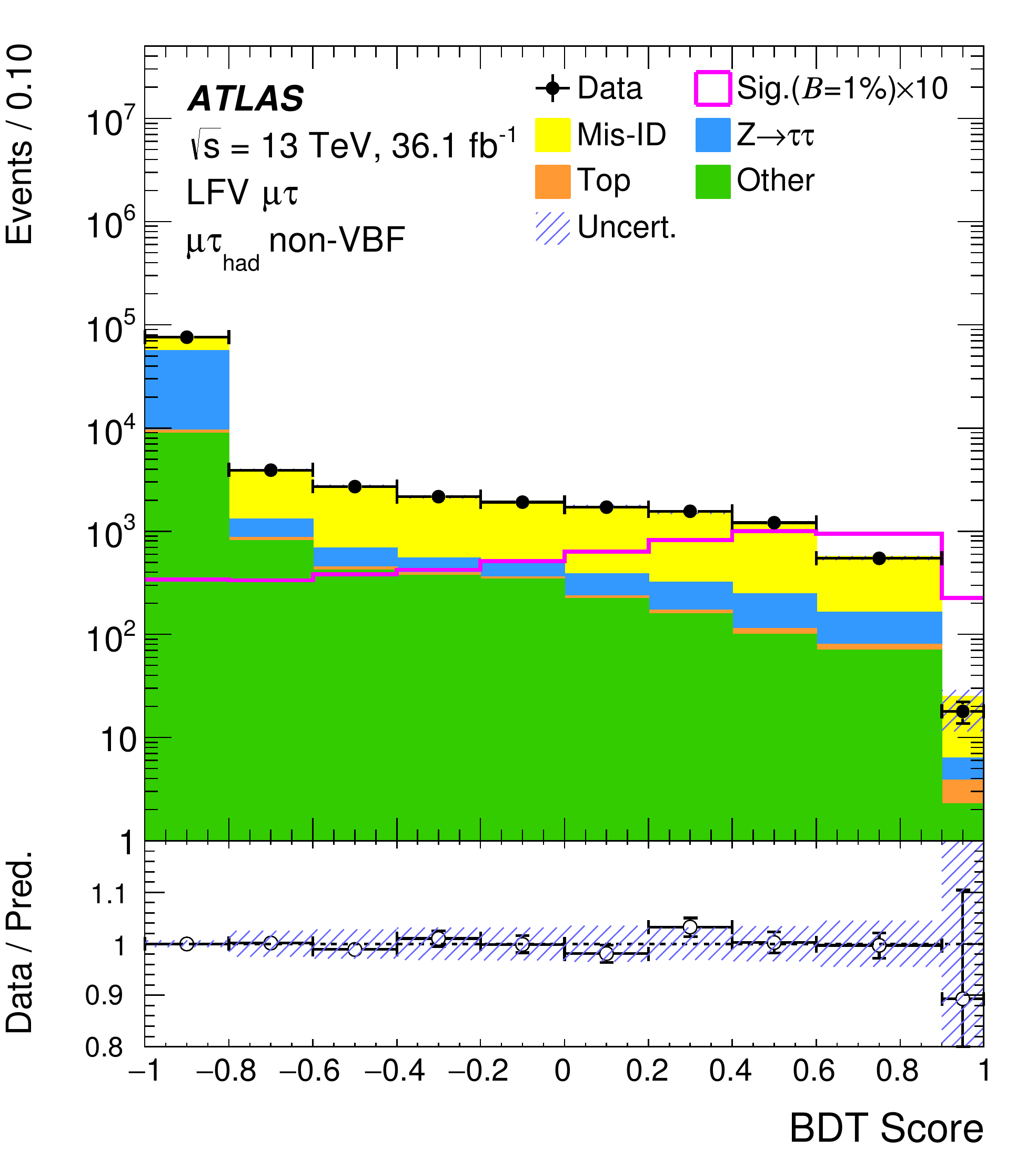}
\includegraphics[width=0.40\textwidth]{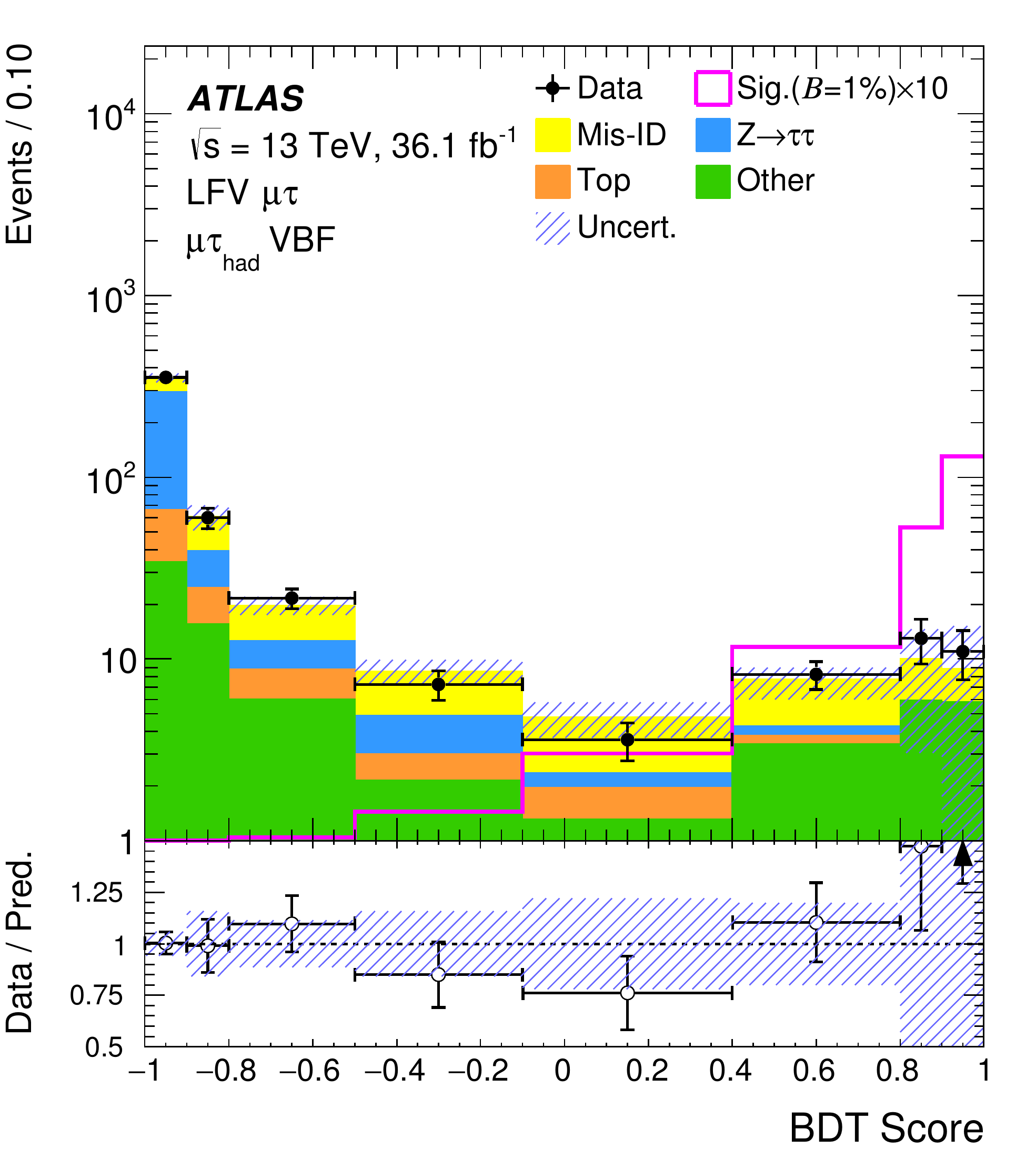}
\caption{Distributions of the BDT score after the background+signal fit in
each signal region of the $\mu\tau$ search, with the LFV signal
overlaid, normalized with $\mathcal{B}(H\to \mu\tau) = 1\%$
and enhanced by a factor $10$ for visibility.
The top and bottom plots display $\mu\tau_{e}$ and $\mu\tauhad$ BDT scores
respectively, the left (right) column corresponds to the non-VBF (VBF)
category. The size of the combined statistical,
experimental and theoretical uncertainties of the background is indicated
by the hatched bands.
The binning is shown as in the statistical analysis.
In the data/background prediction ratio plots, points outside the
displayed $y$-axis range are shown by arrows.
}
\label{fig:lfv_mva_postfit_bdt_mutau}
\end{figure}
 
\begin{table}[htb]
\caption{Summary of the systematic uncertainties and their
impact on the best-fit value of $\mathcal{B}$ in the $H\to e\tau$ and $H\to\mu\tau$
searches. The measured values are obtained by the fit to data,
while the expected values are determined by the fit to a background-only sample.
}
\centering
\small{
\begin{tabular}{lcccc}
\toprule
\multirow{2}{*}{Source of uncertainty}
& \multicolumn{2}{c}{Impact on $\mathcal{B}(H\to e\tau)$ [\%]}
& \multicolumn{2}{c}{Impact on $\mathcal{B}(H\to\mu\tau)$ [\%]}\\
& Measured & Expected & Measured & Expected \\
\midrule
Electron & $+0.05/-0.05$ & $+0.06/-0.06$ & $+0.03/-0.03$ & $+0.02/-0.02$ \\
Muon & $+0.04/-0.04$ & $+0.04/-0.04$ & $+0.10/-0.10$ & $+0.08/-0.10$ \\
\tauhadvis & $+0.02/-0.02$ & $+0.02/-0.02$ & $+0.04/-0.04$ & $+0.04/-0.05$ \\
Jet & $+0.09/-0.08$ & $+0.09/-0.09$ & $+0.11/-0.12$ & $+0.11/-0.12$ \\
$\met$ & $+0.02/-0.02$ & $+0.02/-0.03$ & $+0.05/-0.08$ & $+0.03/-0.05$ \\
$b$-tag & $+0.02/-0.03$ & $+0.03/-0.03$ & $+0.01/-0.01$ & $+0.01/-0.01$ \\
Mis-ID~backg. ($\leplep$) & $+0.08/-0.07$ & $+0.09/-0.08$ & $+0.07/-0.07$ & $+0.07/-0.07$ \\
Mis-ID~backg. ($\lephad$) & $+0.12/-0.11$ & $+0.11/-0.12$ & $+0.11/-0.11$ & $+0.10/-0.10$ \\
Pile-up modelling & $+0.02/-0.01$ & $+0.01/-0.01$ & $+0.05/-0.03$ & $+0.08/-0.06$ \\
Luminosity & $< 0.01$ & $< 0.01$ & $< 0.01$ & $< 0.01$ \\
Background norm. & $+0.05/-0.04$ & $+0.05/-0.03$ & $+0.04/-0.02$ & $+0.05/-0.03$ \\
Theor.~uncert.~(backg.) & $+0.04/-0.03$ & $+0.04/-0.03$ & $+0.08/-0.07$ & $+0.09/-0.09$ \\
Theor.~uncert.~(signal) & $+0.01/-0.01$ & $+0.01/-0.01$ & $+0.04/-0.02$ & $+0.02/-0.02$ \\
MC statistics & $+0.04/-0.04$ & $+0.03/-0.03$ & $+0.04/-0.04$ & $+0.05/-0.04$ \\
 
\midrule
Full systematic & $+0.17/-0.16$ & $+0.17/-0.17$ & $+0.18/-0.18$ & $+0.19/-0.20$ \\
Data statistics & $+0.07/-0.07$ & $+0.07/-0.07$ & $+0.07/-0.07$ & $+0.08/-0.08$ \\
Total & $+0.18/-0.17$ & $+0.18/-0.18$ & $+0.19/-0.19$ & $+0.20/-0.21$ \\
\bottomrule
\end{tabular}
}
\label{tab:NPimpactMVA}
\end{table}
 
\FloatBarrier
 
\section{Results}
The best-fit branching ratios and upper limits are computed while
assuming $\mathcal{B}(H\to\mu\tau) = 0$ for the $H\to e\tau$
search and $\mathcal{B}(H\to e\tau) = 0$ for the $H\to\mu\tau$ search.
The best-fit values of the LFV Higgs boson branching ratios
are equal to
$(0.15^{+0.18}_{-0.17})\%$ and $(-0.22\pm 0.19)\%$ for the $H\to e\tau$
and $H\to\mu\tau$ search,
respectively. In the absence of a significant excess, upper limits
on the LFV branching ratios are set for a Higgs boson with $m_H =
125\,\GeV$. The observed (median expected) 95\% CL upper limits are
$0.47\%$ ($0.34^{+0.13}_{-0.10}\,\%$) and $0.28\%$
($0.37^{+0.14}_{-0.10}\,\%$) for the $H\to e\tau$ and $H\to\mu\tau$ searches,
respectively. These limits are significantly lower than the
corresponding Run~1 limits of Refs.~\cite{HIGG-2014-08,HIGG-2015-09}.
The breakdown of contributions from different signal regions is shown
in Figure~\ref{fig:lfv_mva_limits}.
 
\begin{figure}[htb]
\centering
\includegraphics[width=0.45\textwidth]{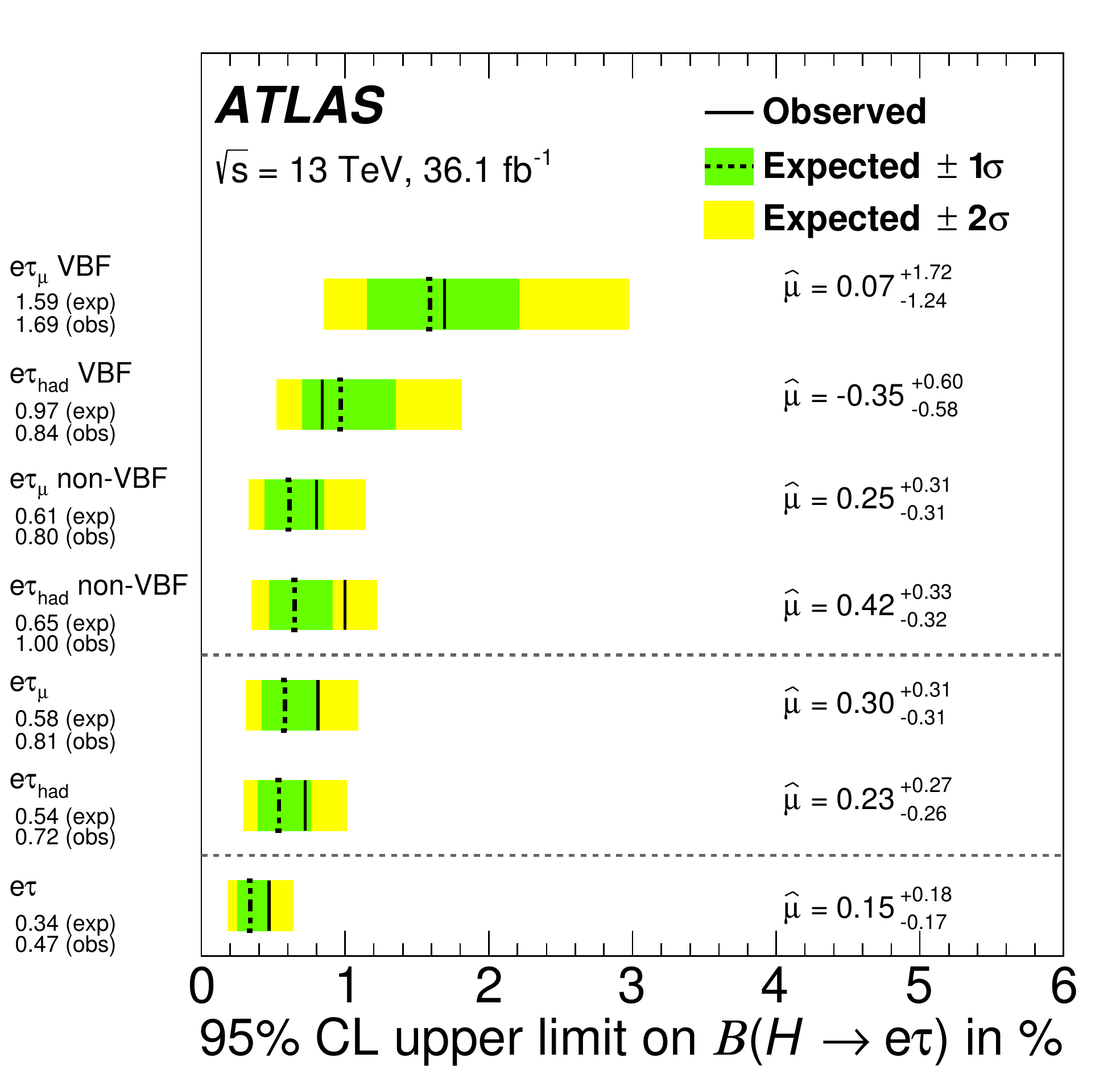}
\includegraphics[width=0.45\textwidth]{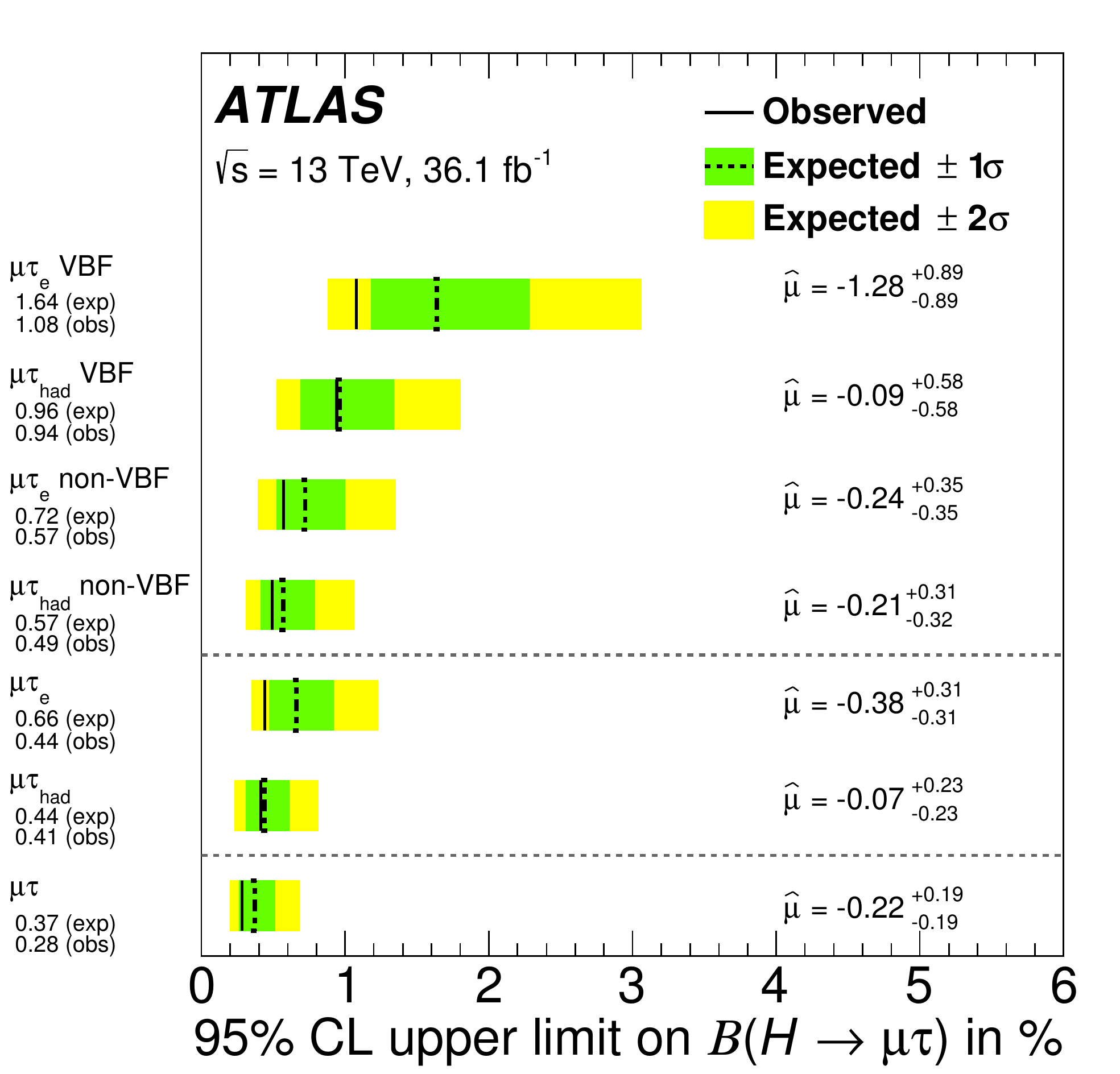}
\caption{Upper limits at 95\% CL on the LFV branching ratios of the
Higgs boson, $H\to e\tau$ (left) and $H\to\mu\tau$ (right), indicated by solid and dashed lines. Best-fit
values of the branching ratios ($\hat{\mu}$) are also given, in \%. The
limits are computed while assuming that either $\mathcal{B}(H\to
\mu\tau)=0$ (left) or $\mathcal{B}(H\to e\tau)=0$ (right).
First, the results of the fits are shown, when only the data of an
individual channel or of an individual category are used; in these cases
the signal and control regions from all other channels/categories are
removed from the fit. These results are finally compared with the full fit
displayed in the last row.}
\label{fig:lfv_mva_limits}
\end{figure}
 
The branching ratio of the LFV Higgs boson decay is related to the
non-diagonal Yukawa coupling matrix elements~\cite{Harnik:2012pb} by the formula
\begin{equation}
|Y_{\ell\tau}|^2+|Y_{\tau\ell}|^2 = \frac{8\pi}{m_H}\,
\frac{\mathcal{B}(H\to\ell\tau)}{1-\mathcal{B}(H\to\ell\tau)}\,\Gamma_{H}(\mathrm{SM}), \notag
\end{equation}
where $\Gamma_{H}(\mathrm{SM}) = 4.07$~MeV~\cite{LHCHiggsCrossSectionWorkingGroup:2013} stands for the Higgs boson width as
predicted by the Standard Model. Thus, the observed limits on the
branching ratio correspond to the following limits on the coupling matrix elements: $\sqrt{|Y_{\tau e}|^2+|Y_{e \tau}|^2} < 0.0020$, and $\sqrt{|Y_{\tau \mu}|^2+|Y_{\mu \tau}|^2} < 0.0015$.
Figure~\ref{fig:couplinglimits} shows the limits on the individual coupling
matrix elements $Y_{\tau \ell}$ and $Y_{\ell \tau}$ together with the
limits from the ATLAS Run~1 analysis and from $\tau\to\ell\gamma$ searches~\cite{Harnik:2012pb,Aubert:2009ag}.
 
\begin{figure}[htb]
\centering
\includegraphics[width=0.45\textwidth]{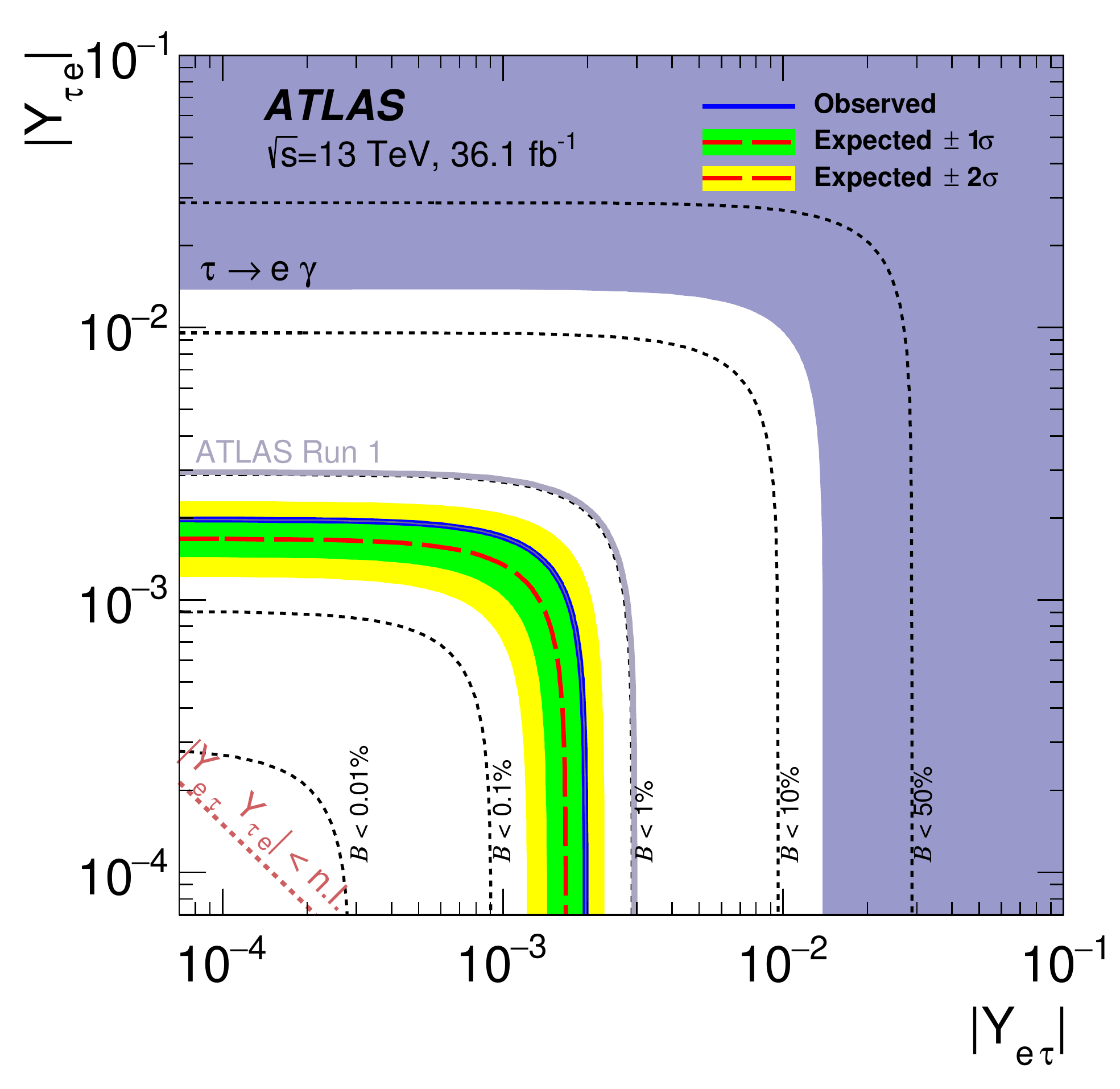}
\includegraphics[width=0.45\textwidth]{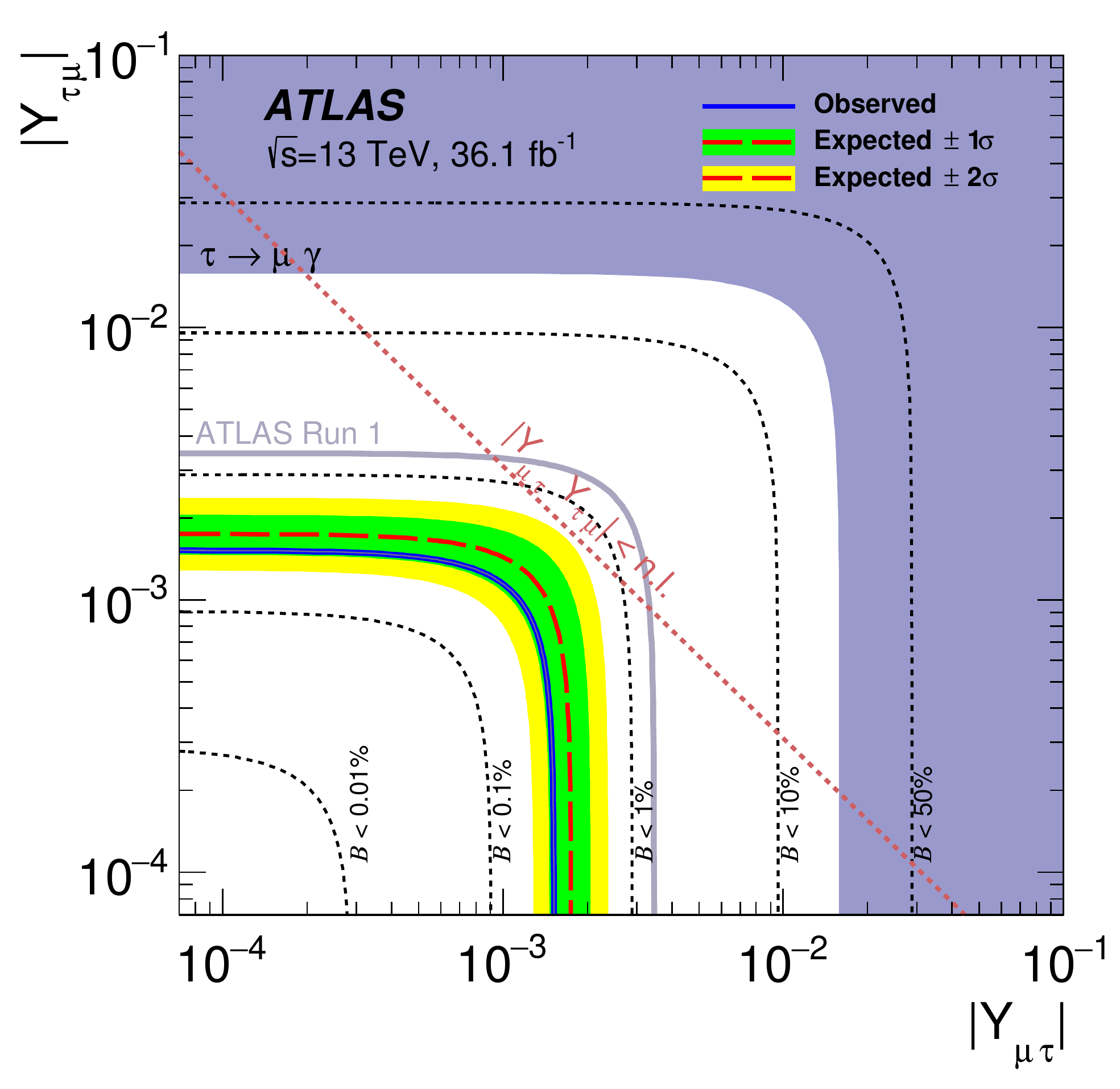}
\caption{Upper limits on the absolute value of the couplings $Y_{\tau \ell}$ and $Y_{\ell \tau}$
together with the limits from the ATLAS Run~1 analysis (light grey line) and the most stringent indirect
limits from $\tau \to \ell \gamma$ searches (dark purple region).
Also indicated are limits corresponding to different branching
ratios ($0.01\%$,
$0.1\%$, $1\%$, $10\%$ and $50\%$) and the naturalness limit
(denoted n.l.) $| Y_{\tau \ell} Y_{\ell \tau} |  \lesssim
\frac{m_\tau m_\ell}{v^2}$~\cite{Harnik:2012pb} where $v$ is the vacuum expectation value of the Higgs field.
}
\label{fig:couplinglimits}
\end{figure}
 
\FloatBarrier
 
\section{Conclusions}
 
Direct searches for the decays $H\to e\tau$ and $H\to\mu\tau$ are performed with proton--proton
collisions recorded by the ATLAS detector at the LHC corresponding to
an integrated luminosity of $36.1\,\ifb$ at a centre-of-mass energy
of $\sqrt{s} = 13\,\TeV$. No significant excess is observed above the expected background from Standard Model processes.
The observed (expected) upper limits at 95\% confidence level on the branching ratios of
$H\to e\tau$ and $H\to\mu\tau$ are
$0.47\%$ ($0.34^{+0.13}_{-0.10}\,\%$) and $0.28\%$
($0.37^{+0.14}_{-0.10}\,\%$), respectively.
These limits are more stringent by a factor of 2 (5)
than the corresponding limits for the $H\to e\tau$ ($H\to\mu\tau$) decay
determined by ATLAS at $\sqrt{s} = 8$ TeV.
\FloatBarrier
 
\section*{Acknowledgements}

 
We thank CERN for the very successful operation of the LHC, as well as the
support staff from our institutions without whom ATLAS could not be
operated efficiently.
 
We acknowledge the support of ANPCyT, Argentina; YerPhI, Armenia; ARC, Australia; BMWFW and FWF, Austria; ANAS, Azerbaijan; SSTC, Belarus; CNPq and FAPESP, Brazil; NSERC, NRC and CFI, Canada; CERN; CONICYT, Chile; CAS, MOST and NSFC, China; COLCIENCIAS, Colombia; MSMT CR, MPO CR and VSC CR, Czech Republic; DNRF and DNSRC, Denmark; IN2P3-CNRS, CEA-DRF/IRFU, France; SRNSFG, Georgia; BMBF, HGF, and MPG, Germany; GSRT, Greece; RGC, Hong Kong SAR, China; ISF and Benoziyo Center, Israel; INFN, Italy; MEXT and JSPS, Japan; CNRST, Morocco; NWO, Netherlands; RCN, Norway; MNiSW and NCN, Poland; FCT, Portugal; MNE/IFA, Romania; MES of Russia and NRC KI, Russian Federation; JINR; MESTD, Serbia; MSSR, Slovakia; ARRS and MIZ\v{S}, Slovenia; DST/NRF, South Africa; MINECO, Spain; SRC and Wallenberg Foundation, Sweden; SERI, SNSF and Cantons of Bern and Geneva, Switzerland; MOST, Taiwan; TAEK, Turkey; STFC, United Kingdom; DOE and NSF, United States of America. In addition, individual groups and members have received support from BCKDF, CANARIE, CRC and Compute Canada, Canada; COST, ERC, ERDF, Horizon 2020, and Marie Sk{\l}odowska-Curie Actions, European Union; Investissements d' Avenir Labex and Idex, ANR, France; DFG and AvH Foundation, Germany; Herakleitos, Thales and Aristeia programmes co-financed by EU-ESF and the Greek NSRF, Greece; BSF-NSF and GIF, Israel; CERCA Programme Generalitat de Catalunya, Spain; The Royal Society and Leverhulme Trust, United Kingdom.
 
The crucial computing support from all WLCG partners is acknowledged gratefully, in particular from CERN, the ATLAS Tier-1 facilities at TRIUMF (Canada), NDGF (Denmark, Norway, Sweden), CC-IN2P3 (France), KIT/GridKA (Germany), INFN-CNAF (Italy), NL-T1 (Netherlands), PIC (Spain), ASGC (Taiwan), RAL (UK) and BNL (USA), the Tier-2 facilities worldwide and large non-WLCG resource providers. Major contributors of computing resources are listed in Ref.~\cite{ATL-GEN-PUB-2016-002}.
 

\printbibliography
\clearpage
 
\begin{flushleft}
{\Large The ATLAS Collaboration}

\bigskip

G.~Aad$^\textrm{\scriptsize 101}$,    
B.~Abbott$^\textrm{\scriptsize 128}$,    
D.C.~Abbott$^\textrm{\scriptsize 102}$,    
A.~Abed~Abud$^\textrm{\scriptsize 70a,70b}$,    
K.~Abeling$^\textrm{\scriptsize 53}$,    
D.K.~Abhayasinghe$^\textrm{\scriptsize 93}$,    
S.H.~Abidi$^\textrm{\scriptsize 167}$,    
O.S.~AbouZeid$^\textrm{\scriptsize 40}$,    
N.L.~Abraham$^\textrm{\scriptsize 156}$,    
H.~Abramowicz$^\textrm{\scriptsize 161}$
% ...
\bigskip
\\

% ...
$^{40}$Niels Bohr Institute, University of Copenhagen, Copenhagen; Denmark.\\
% ...
$^{53}$II. Physikalisches Institut, Georg-August-Universit\"{a}t G\"ottingen, G\"ottingen; Germany.\\
% ...
$^{70}$$^{(a)}$INFN Sezione di Pavia;$^{(b)}$Dipartimento di Fisica, Universit\`a di Pavia, Pavia; Italy.\\
% ...
$^{93}$Department of Physics, Royal Holloway University of London, Egham; United Kingdom.\\
% ...
$^{101}$CPPM, Aix-Marseille Universit\'e, CNRS/IN2P3, Marseille; France.\\
$^{102}$Department of Physics, University of Massachusetts, Amherst MA; United States of America.\\
% ...
$^{128}$Homer L. Dodge Department of Physics and Astronomy, University of Oklahoma, Norman OK; United States of America.\\
% ...
$^{156}$Department of Physics and Astronomy, University of Sussex, Brighton; United Kingdom.\\
% ...
$^{161}$Raymond and Beverly Sackler School of Physics and Astronomy, Tel Aviv University, Tel Aviv; Israel.\\
% ...
$^{167}$Department of Physics, University of Toronto, Toronto ON; Canada.\\
% ...
$^{a}$ Also at CERN, Geneva; Switzerland.\\
$^{b}$ Also at CPPM, Aix-Marseille Universit\'e, CNRS/IN2P3, Marseille; France.\\
% ...

\end{flushleft}

 
\end{document}